\documentclass[prd,aps,preprintnumbers,showpacs,twocolumn,nofootinbib,superscriptaddress,notitlepage,floatfix]{revtex4-1}

\usepackage{ulem}
\usepackage{amsmath,amssymb,amsfonts,mathtools}
\usepackage{bm}
\usepackage{graphicx}
\usepackage{booktabs}
\usepackage{array}[=2016-10-06]
\usepackage{placeins}
\usepackage{slashed}
\usepackage{multirow}
\usepackage{xcolor}
\usepackage{graphicx}
\usepackage[caption=false]{subfig}
\usepackage[hidelinks]{hyperref}

\allowdisplaybreaks[4]

\newcommand{\dd}{\mathrm{d}}
\newcommand{\order}{\mathcal{O}}
\newcommand{\as}{\alpha_s}
\newcommand{\LQCD}{\Lambda_{\mathrm{QCD}}}
\newcommand{\Ltag}{\Lambda_{\mathrm{tag}}}
\newcommand{\Mtag}{\mathcal{M}}
\newcommand{\Simp}{\mathcal{S}}
\newcommand{\CpA}{C_p^A}
\newcommand{\RgPA}{R_g^{p/A}}
\newcommand{\RgSRC}{R_g^{\mathrm{SRC}\,N}}
\newcommand{\nn}{\nonumber}

\begin{document}

\title{Short-Range Correlation Scaling of Nuclear Parton Modifications: \\
Consistency Tests and Recoil-Tagged Charm-Production DIS}

\author{Yong Zheng}
\email{yzheng2018@lzu.edu.cn}
\affiliation{Frontiers Science Center for RareIsotopes, and School of Nuclear Science and Technology, Lanzhou University, Lanzhou 730000, China}

\author{Shu-Man Hu}
\email{hushm2025@lzu.edu.cn, corresponding author}
\affiliation{Frontiers Science Center for RareIsotopes, and School of Nuclear Science and Technology, Lanzhou University, Lanzhou 730000, China}

\author{Wei Wang}
\email{wei.wang@sjtu.edu.cn, corresponding author}
\affiliation{State Key Laboratory of Dark Matter Physics, Shanghai Key Laboratory for Particle Physics and Cosmology, Key Laboratory for Particle Astrophysics and Cosmology (MOE), School of Physics and Astronomy, Shanghai Jiao Tong University, Shanghai 200240, China}
\affiliation{Southern Center for Nuclear-Science Theory (SCNT), Institute of Modern Physics, Chinese Academy of Sciences, Huizhou 516000, Guangdong Province, China}

\author{Ji Xu}
\email{xuji@lzu.edu.cn, corresponding author}
\affiliation{Frontiers Science Center for RareIsotopes, and School of Nuclear Science and Technology, Lanzhou University, Lanzhou 730000, China}

\author{Shuai Zhao}
\email{zhaos@tju.edu.cn, corresponding author}
\affiliation{Department of Physics and Center for Joint Quantum Studies, School of Science, Tianjin University, Tianjin 300350, China}

\begin{abstract}
We investigate short-range correlation (SRC) scaling of nuclear gluon modifications using nCTEQ SRC abundances and EPPS21 gluon distributions. Rescaling the EPPS21 central modifications by these abundances yields an approximately common curve for six nuclei from carbon to lead over $0.1\leq x\leq0.4$ at $Q^2=10$ and $100\,\mathrm{GeV}^2$. The SRC interpretation is supported by the broad agreement between the fitted abundances and high-momentum proton fractions from variational Monte Carlo calculations. Using an SRC proton fraction of approximately $20\%$ in carbon, the measured EMC–SRC correlation also gives a pure-SRC EMC slope of $1.94\pm0.12$, consistent with the nCTEQ estimate $2.02\pm0.20$. To test the connection between this scaling and SRC configurations, we propose recoil-tagged charm-production in deuteron deep-inelastic scattering (DIS), combining a backward high-momentum proton tag with a gluon-sensitive hard process. We formulate leading-power collinear factorization through an extended fracture function, which reduces within the plane-wave impulse approximation to a deuteron light-front spectral factor times a recoil-conditioned neutron gluon distribution. The tagged charm reduced cross section and its ratio to inclusive tagged DIS provide complementary observables for comparing gluon modifications in SRC-enriched configurations with the scaling inferred from inclusive nPDFs.
\end{abstract}

\maketitle

\section{Introduction}
\label{sec:introduction}

Understanding how quarks and gluons share momentum in nucleons and nuclei is a central problem in quantum chromodynamics (QCD).、
At a fixed factorization scale, parton distribution functions (PDFs) describe the nonperturbative structure needed for high-energy scattering processes \cite{ParticleDataGroup:2026mpi,Gao:2017yyd,Hou:2019efy,NNPDF:2021njg,Ball:2017otu}. 
Nuclear PDFs (nPDFs) extend this description to bound nucleons \cite{Eskola2022EPPS21,AbdulKhalek:2022fyi,Duwentaster:2022kpv,Helenius:2021tof,Denniston2024SRCnPDF}. 
Comparing PDFs and nPDFs is particularly valuable because nucleons and nuclei represent two successive levels of matter governed by the same underlying QCD dynamics, with quarks and gluons confined within nucleons and residual strong interactions binding these nucleons into nuclei.
Their difference therefore provides a unique window into the role of the nuclear many-body environment in partonic structure.

Short-range correlation (SRC) nucleon pairs are especially important in this context because they involve small internucleon separations, large relative momentum, and high nucleon virtuality, where nuclear interaction is strongest and modifications of partonic structure may be enhanced \cite{Frankfurt:1993sp,Frankfurt:1988nt,Arrington:2011xs,Miller:2020eyc,Schmidt2020Core}. 
The empirical correlation between the EMC effect \cite{EuropeanMuon:1983wih,EuropeanMuon:1988lbf,EuropeanMuon:1988tpw,NewMuonNMC:1990xyw,Seely:2009gt} and SRC abundance suggests that SRCs may provide a promising link between low-energy nuclear dynamics and high-energy partonic structure \cite{Hen:2016kwk,Schmookler2019Modified,Subedi2008ColdDense,Weinstein:2010rt,CLAS:2005ola,Hen:2012fm,Hen:2014nza,CLAS:2018yvt,Hatta:2019ocp,Yang:2023zmr,Huang:2021cac,Chen2017SRCEMC,Wang:2025pcu,Yan:2026blw}.
This correlation does not by itself establish causality, but the underlying hypothesis should be tested in different ways such as global fitting and direct experimental measurements.

Conventional global analyses parametrize the effective PDFs of bound nucleons and constrain them with inclusive nuclear data.
The resulting distributions are averages over the nuclear wave function and do not identify which configurations generate the observed modifications.  
This leaves room for SRC-based, virtuality, convolution, and other nuclear mechanisms \cite{CruzTorres2021Factorization,Peng2026PartonSRC,Wang2020SRCEMC,Frankfurt:1994hf,Dutta:2012ii}.
Recently, the nCTEQ analysis made the SRC picture quantitative within a global nPDF fit \cite{Denniston2024SRCnPDF}. 
It decomposed each bound nucleon distribution into a free mean-field component and a universal SRC component, with nucleus dependent coefficients $C_p^A$ and $C_n^A$, and achieved a fit quality comparable to, or better than, that of conventional parametrizations. 
This construction connects partonic distributions with quantities that have a nuclear structure interpretation.
However, inclusive data constrain only the product of the SRC abundance and the modification carried by an SRC nucleon.
As pointed out in Ref.\,\cite{Paakkinen2026Scaling}, a common rescaling of $C_N^A$, accompanied by the corresponding inverse rescaling of the SRC modification, leaves effective nPDFs unchanged.

In this work, we first compare the SRC-abundance coefficients $C^A_p$ extracted by nCTEQ with high-momentum proton fractions obtained from variational Monte Carlo (VMC) calculations \cite{Wiringa2014Momentum}.
We then combine the experimental $a_2^p$-$|\dd R_{\mathrm{EMC}}/\dd x|$ relation with a measured SRC fraction for \(^{12}\mathrm C\) to estimate the EMC slope of a hypothetical ``pure'' SRC protons \cite{Schmookler2019Modified,Subedi2008ColdDense}. 
Together, these comparisons test the nuclear dependence of SRC abundance and the characteristic size of its quark modification in the EMC region, although they do not establish SRCs as the unique source of partonic modifications.

We next examine whether the SRC scaling extends to gluon nuclear modifications.
The nCTEQ analysis parametrized SRC contributions to both quark and gluon nPDFs, with gluon sensitivity arising mainly from deep-inelastic scattering (DIS) scaling violations and $W/Z$ production.
To test this picture across global analyses, we combine the nCTEQ abundance coefficients $C_p^A$ with the EPPS21 gluon nuclear modifications \cite{Eskola2022EPPS21}.
This allows us to reconstruct an SRC gluon ratio for each nucleus within the two-component ansatz.
If the proposed scaling holds, the reconstructed ratios should coincide at fixed $(x,Q^2)$.

The inclusive studies above motivate a test of SRC scaling at the level of selected nuclear configurations.
To this end, we consider recoil-tagged charm-production in deuteron DIS.
A backward high-momentum proton selects an SRC-enriched sample and, within the plane-wave impulse approximation (PWIA), constrains the momentum and virtuality of the active neutron~\cite{Melnitchouk1997Tagged,StrikmanWeiss2018,Hauenstein2024LAD}.
Inclusive charm-production provides gluon sensitivity through photon-gluon fusion at leading order.
Together, these measurements connect a gluon-sensitive observable with a selected nuclear configuration.

Starting from the exact tagged hadronic tensor, we express the tagged structure functions at leading power as convolutions of massive hard coefficients with a recoil-differential extended fracture function~\cite{TrentadueVeneziano1994,Grazzini1998CutVertices,Collins1998Diffractive}.
Within PWIA, we match this function onto the product of a deuteron light-front spectral factor and a recoil-conditioned neutron gluon distribution.
This framework provides a basis for extracting the conditional gluon distribution and testing the SRC interpretation of the inclusive scaling.

The paper is organized as follows. 
Sec.\,\ref{sec:src-framework} introduces the SRC-based nPDF parametrization and its rescaling freedom. 
Sec.\,\ref{sec:independent-tests} compares the fitted SRC-abundance coefficient $\CpA$ with VMC high-momentum proton fractions $C_p^{A,\rm VMC}$ and examines the EMC slope of pure-SRC based on the measured EMC-SRC correlation. 
Sec.\,\ref{sec:gluon} reconstructs the candidate SRC gluon ratio from EPPS21 and performs the rank-one test. 
Sec.\,\ref{sec:tagged-charm} presents the factorization and PWIA framework for recoil-tagged charm-production in DIS and defines the proposed observables. 
Sec.\,\ref{sec:summary} summarizes our results and discusses future prospects. Apps.\,\ref{app:kinematics}-\ref{app:coefficient} collect the recoil kinematics, tensor contraction, and massive photon-gluon coefficient.

\section{SRC-Based Nuclear PDFs and the Rescaling Invariance}
\label{sec:src-framework}

In this section, we begin by reviewing the two-component parametrization used in the nCTEQ analysis \cite{Denniston2024SRCnPDF}.
For a proton bound in a nucleus $A$, the parametrization reads
\begin{eqnarray}\label{eq:two-component-proton}
 f_i^{p/A}(x,Q^2)
 =\left(1-\CpA\right)f_i^p(x,Q^2)
 +\CpA f_i^{\mathrm{SRC}p}(x,Q^2) \,,\nn\\
\end{eqnarray}
where $f_i^{p/A}$ is the PDF of an average bound proton, $f_i^p$ the free-proton PDF, and $f_i^{\mathrm{SRC}p}$ a nucleus-indenpendent PDF assigned to a proton in an SRC pairs. 
The parton label $i$ includes quarks, antiquarks, and gluons.
The corresponding per-nucleon distribution of the full nucleus is
\begin{equation}\label{eq:full-nucleus-pdf}
 f_i^A(x,Q^2) = \frac{Z}{A} \, f_i^{p/A}(x,Q^2) + \frac{N}{A} \, f_i^{n/A}(x,Q^2) \,,
\end{equation}
with the mass number $A=Z+N$. 
We use the same two-component form for $f_i^{n/A}$ as for $f_i^{p/A}$.
Ref.~\cite{Denniston2024SRCnPDF} implicitly assumes that the proton and neutron distributions are related by isospin.
In particular, the proton and neutron gluon distributions are taken to be equal to each other.

It is convenient to rewrite Eq.\,\eqref{eq:two-component-proton} as a modification ratio. 
For any flavor with a nonzero free proton distribution, define
\begin{equation}\label{eq:src-ratio-definition}
\begin{cases}
 R_i^{p/A}(x,Q^2) \equiv f_i^{p/A}(x,Q^2)/f_i^p(x,Q^2) \,,\\[10pt]
 R_i^{\mathrm{SRC}p}(x,Q^2) \equiv f_i^{\mathrm{SRC}p}(x,Q^2)/f_i^p(x,Q^2) \,.
\end{cases}
\end{equation}
The two-component ansatz then becomes
\begin{equation}\label{eq:ratio-two-component}
 R_i^{p/A}(x,Q^2)-1 = C_p^A \left(R_i^{\mathrm{SRC}p}(x,Q^2) -1 \right) \,.
\end{equation}
Eq.\,\eqref{eq:ratio-two-component} shows that the nuclear modification, rather than the full ratio $R_i^{p/A}$, factorizes into a nucleus-dependent coefficient and a common $x$-dependent function. 
At fixed $Q^2$, we can construct a matrix with elements $R_i^{p/A_\ell}(x_m)-1$, where the rows label nuclei $A_\ell$ and the columns label $x_m$. 
This matrix should be rank one if the two-component ansatz is valid. 
Here, subtracting unity is essential because all of the full ratios $R_i^{p/A}$ for different nuclei contain the same free nucleon baseline. 
Since this common baseline can dominate over the nuclear modifications, a rank-one test applied directly to $R_i^{p/A}$ may appear successful even when the modifications $R_i^{p/A}-1$ do not share a common shape.
We therefore also conduct the rank-one test after the free nucleon baseline has been removed.

Within the two-component ansatz, $\CpA$ is interpreted as the probability that a proton in nucleus $A$ belongs to an SRC pair. 
This picture is natural because correlated nucleons dominate high-momentum nuclear configurations \cite{Schmookler2019Modified}.
However, inclusive data alone cannot determine $C^A_p$ as an absolute probability.
Such data determine only the average bound proton PDF, and the same average can be obtained by changing $C_p^A$ while compensating for it with a change in the SRC proton PDF $f_i^{\mathrm{SRC}p}$. 
Therefore, identifying $C_p^A$ as an absolute SRC probability requires additional information.

To make this point more evident, define the modification carried by the SRC component,
\begin{equation} \label{eq:src-modification}
 \delta f_i^{\mathrm{SRC}p}(x,Q^2)
 \equiv f_i^{\mathrm{SRC}p}(x,Q^2)-f_i^p(x,Q^2) \,.
\end{equation}
The bound-nucleon modification then depends on the product
\begin{equation} \label{eq:observable-product}
 f_i^{p/A}-f_i^p = C_p^A\,\delta f_i^{\mathrm{SRC}p} \,.
\end{equation}
Consequently, the global transformation
\begin{equation} \label{eq:rescaling-invariance}
 C_N^A\longrightarrow \lambda C_N^A \,,
 \qquad
 \delta f_i^{\mathrm{SRC}N}\longrightarrow
 \frac{1}{\lambda}\delta f_i^{\mathrm{SRC}N} \,,
\end{equation}
with the same constant $\lambda$ for all nuclei, leaves all physical nuclear PDFs unchanged, as pointed out in Ref.\,\cite{Paakkinen2026Scaling}.
It also preserves all observables and sum rules for the physical bound nucleon PDFs. 
Bounds such as $0\leq C_N^A\leq1$ restrict the allowed interval of $\lambda$, but they generally do not select a unique point. 
Inclusive data can determine the products in Eq.\,\eqref{eq:observable-product}, as well as the relative nuclear dependence $C_p^{A_1}/C_p^{A_2}$, and they cannot separately determine the absolute value of $C_p^A$ and the explicit strength of $\delta f_i^{\mathrm{SRC}p}$.

For two nuclei at the same $(x,Q^2)$, Eq.\,\eqref{eq:ratio-two-component} implies
\begin{equation} \label{eq:relative-nuclear-scaling}
 \frac{R_i^{p/A_1}(x,Q^2)-1}{R_i^{p/A_2}(x,Q^2)-1}
 =\frac{C_p^{A_1}}{C_p^{A_2}} \,,
\end{equation}
whenever the denominator is nonzero. This relation is invariant under Eq.\,\eqref{eq:rescaling-invariance}. Thus the relative $A$ dependence is a genuine prediction of the parametrization even before its absolute normalization is fixed. By contrast, a statement about absolute value and modification must therefore identify the external information which chooses one normalization condition.

The rescaling freedom nevertheless leaves several testable relations, which address the SRC interpretation in successive steps. First, the nuclear dependence of \(C_p^A\) can be compared with high-momentum proton fractions from VMC calculations, testing its interpretation as an SRC abundance. Second, an external SRC normalization can be combined with the measured EMC--SRC correlation to test the quark modification inferred by nCTEQ. Third, the same \(C_p^A\) coefficients can be used to determine whether the gluon modifications from EPPS21 reduce to a common \(x\)-dependent shape. We examine the first two comparisons in Sec.\,\ref{sec:independent-tests} and the third in Sec.\,\ref{sec:gluon}. Agreement in all three tests would provide stronger support for the SRC interpretation than any single comparison. However, none of these tests directly selects an SRC configuration in the measured scattering event. They therefore cannot establish that the observed partonic modification occurs in an SRC pair, because the correlated partner nucleon is not detected event by event. A more direct test therefore requires selecting SRC-enriched configurations, as proposed through recoil-tagged DIS in Sec.\,\ref{sec:tagged-charm}.

%%%%%%%%%%%%%%%%%%%%%%%%%%%%%%%%%%%%%%%%%%%%%%%%%%%%%%%%%%%%%%%%%%%%%%%%%%%%%%%%%%%%%%%%%%%%%%%%%%%%%%%%%%%%%%%%%%%%%%%%%%%%%%%%%%%%%%%%%%%%%%%%%%%%%%%%%%%%%%%%%%%%%%%%%%%%%%%%%%%%%%%%%%%%%%%%%%%%%%%%%%%%%%%%%%%%%%%
\section{Independent Nuclear Structure and Quark Sector Tests}
\label{sec:independent-tests}

We first examine the two ingredients of the SRC-based parametrization that can be confronted with information from other studies \cite{Schmookler2019Modified,Subedi2008ColdDense,Wiringa2014Momentum}.
The first comparison concerns the magnitude and nuclear dependence of the abundance coefficient $\CpA$.
The second concerns the characteristic size of the quark sector modification.

\subsection{Abundance test: VMC momentum distributions versus fitted \texorpdfstring{\(C_p^A\)}{CpA} in nCTEQ}
\label{sec:vmc-test}

VMC calculations provide a low-energy nuclear structure benchmark for the fitted abundance in nCTEQ. 
We use the spin-summed ground-state proton momentum distributions $\rho_p^A(k)$ calculated with the AV18 two-nucleon and UX three-nucleon interactions in Ref.\,\cite{Wiringa2014Momentum}.
These wave functions and momentum distributions have been widely used in nuclear physics studies \cite{Carlson:2014vla,CiofidegliAtti:2015lcu,LabHallA:2014wqo}.
We normalize the spin-summed proton momentum distribution such that
\begin{equation} \label{eq:vmc-normalization}
 \int\frac{\dd^3\bm k}{(2\pi)^3}\,\rho_p^A(\bm k) = Z \,.
\end{equation}
We define the fraction of protons above a momentum threshold \(k_{\mathrm{cut}}\) by
\begin{equation} \label{eq:vmc-fraction}
 C_p^{A,\mathrm{VMC}}(k_{\mathrm{cut}})
 =\frac{
 \displaystyle\int_{|\bm k|>k_{\mathrm{cut}}}
 \frac{\dd^3\bm k}{(2\pi)^3}\rho_p^A(\bm k)}{
 \displaystyle\int
 \frac{\dd^3\bm k}{(2\pi)^3}\rho_p^A(\bm k)} \,.
\end{equation}
The denominator equals $Z$. The onset of the SRC-dominated tail has no unique, model-independent momentum. We therefore use \(k_{\mathrm{cut}}=280\) and \(300\,\mathrm{MeV}\).  These values span a typical boundary between mean-field and high-momentum configurations \cite{Wang2020SRCEMC}. 
The result depends on both the cutoff and the nuclear Hamiltonian.
Here, we compare its size and nuclear dependence with the fitted $\CpA$.

The VMC fraction is defined by a momentum cut on the one-body proton distribution, whereas \(C_p^A\) is fitted to inclusive partonic data without such a cut. High-momentum protons are expected to be enriched in SRC configurations, so agreement in magnitude and nuclear dependence provides a consistency test of the SRC interpretation of $C_p^A$. We normalize the VMC high-momentum yield by \(Z\), matching the proton-specific coefficient \(C_p^A\) in nCTEQ. This choice is particularly relevant for asymmetric nuclei, where the high-momentum proton fraction can differ from the fraction averaged over all nucleons.

\begin{figure*}[t]
\centering
\includegraphics[width=0.43\textwidth]{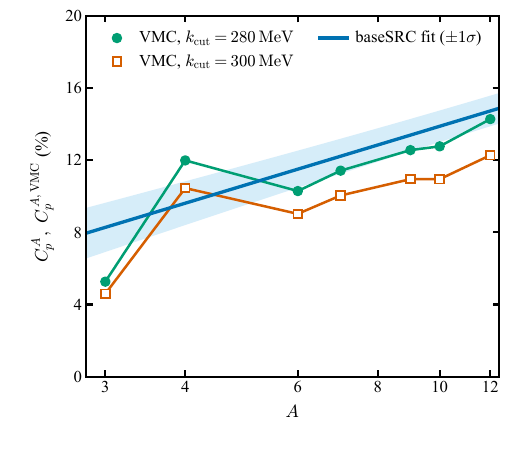}
\caption{VMC high-momentum proton fractions compared with the fitted nCTEQ abundance. The blue curve and band show $C_p^A$ in Eq.\,\eqref{eq:cpa-log-fit}; green circles and orange open squares correspond to $k_{\mathrm{cut}}=280$ and $300\,\mathrm{MeV}$ for $C_p^{A,\mathrm{VMC}}$, respectively. The mass number axis is logarithmic.}
\label{fig:vmc}
\end{figure*}

The nCTEQ analysis displays a logarithmic fit in $A$ for the central SRC proton abundance. Reconstructing this fit from the central values presented in Ref.\,\cite{Denniston2024SRCnPDF} gives:
\begin{equation} \label{eq:cpa-log-fit}
\,\CpA(\%) = (3.16\pm2.02)+(10.71\pm1.60)\log_{10}A \,.
\end{equation}
The uncertainties follow from a diagonal-error fit. Eqs.\,(\ref{eq:vmc-fraction}) and (\ref{eq:cpa-log-fit}) can then be compared.
Tab.\,\ref{tab:vmc} and Fig.\,\ref{fig:vmc} show the comparison.
The VMC results obtained with both choices of $k_{\rm cut}$ reproduce the general increase of $\CpA$ with $A$ and are similar in magnitude to the nCTEQ values. 
Overall, the results with $k_{\rm cut}=280\,\mathrm{MeV}$ show better agreement with the nCTEQ fit.
Although $C_p^{A,\mathrm{VMC}}$ and $\CpA$ follow a similar nuclear dependence, visible differences remain. 
Such differences may be caused by the VMC calculation retaining the shell structure of each nucleus and by the gradual transition from mean-field to correlated nucleon motion, which makes it difficult to determine $k_{\rm cut}$ accurately.

\begin{table*}[t]
\caption{VMC high-momentum proton fractions and the fitted nCTEQ coefficient.  All dimensionless fractions are displayed in percent.}
\label{tab:vmc}
\begin{ruledtabular}
\begin{tabular}{ccccc}
Nucleus & \(A\) & $\CpA$ & $C_p^{A,\mathrm{VMC}}(280\,\mathrm{MeV})$ & $C_p^{A,\mathrm{VMC}}(300\,\mathrm{MeV})$ \\[2pt]\hline\\[-5pt]
\(^{3}\mathrm{He}\)  & 3  & \(8.27\pm1.36\)  & 5.27  & 4.60\\
\(^{4}\mathrm{He}\)  & 4  & \(9.61\pm1.21\)  & 11.98 & 10.45\\
\(^{6}\mathrm{Li}\)  & 6  & \(11.50\pm1.03\) & 10.29 & 9.02\\
\(^{7}\mathrm{Li}\)  & 7  & \(12.22\pm0.97\) & 11.42 & 10.04\\
\(^{9}\mathrm{Be}\)  & 9  & \(13.38\pm0.90\) & 12.55 & 10.95\\
\(^{10}\mathrm{B}\)  & 10 & \(13.87\pm0.87\) & 12.76 & 10.94\\
\(^{12}\mathrm{C}\) & 12 & \(14.72\pm0.85\) & 14.27 & 12.27\\
\end{tabular}
\end{ruledtabular}
\end{table*}

The agreement shown in Fig.\,\ref{fig:vmc} supports the interpretation of the fitted $\CpA$ as an SRC abundance.
More specifically, it shows that the relative \(A\) dependence inferred from high-energy inclusive data is compatible with the high-momentum content of realistic nuclear wave functions.
It should note that the results in Fig.\,\ref{fig:vmc} is not a precision goodness-of-fit test. 
The main point is: two analyses at very different scales find a similar trend for different nuclei without using each other as input, which implies the validity of factorization (separating $\CpA$ from $f_i^{\mathrm{SRC}p}$).

\subsection{Shape test: reconstruction of the pure-SRC EMC slope}
\label{sec:emc-test}

We next test the quark modification part of the two-component picture.
The experimental EMC-SRC correlation relates the magnitude of the per-proton EMC slope to the per-proton quasielastic scaling coefficient $a_2^p$ \cite{Hen:2016kwk,Weinstein:2010rt,Schmookler2019Modified}.
We use the deuteron-anchored form, in which the linear relation is constrained to pass through the deuteron reference point $a_2^p=1$ and $\left|\dd R_{\mathrm{EMC}}^p/\dd x\right|=0$,
\begin{equation} \label{eq:emc-src-linear}
 \left|\frac{\dd R_{\mathrm{EMC}}^p}{\dd x}\right|
 =C_{\mathrm{slope}}\left(a_2^p-1\right) \,.
\end{equation}
The absolute value fixes the sign convention for the vertical axis.  
A fit to the measured points gives
\begin{equation}\label{eq:emc-slope-coefficient}
 C_{\mathrm{slope}}=0.0906\pm0.0040 \,.
\end{equation}
The two axes in Eq.\,\eqref{eq:emc-src-linear} come from different measurements.
The quantity $a_2^p$ is obtained from the quasielastic plateau at large Bjorken $x$. 
The EMC slope is measured in the DIS process. 
Their linear relation is therefore not a kinematic identity. 
It links a low-energy property of the nuclear wave function to a high-energy change in quark sensitive structure functions. 
For this reason, the correlation provides information independent of an nPDF fit.

We define a hypothetical pure-SRC proton ensemble by extrapolating the SRC-proton fraction to unity, thereby removing the contribution from protons outside SRC configurations.
To perform an independent test, we use the experimental measurement that about $20\%$ of the nucleons in $^{12}\mathrm C$ belong to SRC pairs \cite{Subedi2008ColdDense}.
If the measured signal scales linearly with this fraction, then, using $a_2^p(^{12}\mathrm C)=4.49\pm0.17$ \cite{Schmookler2019Modified}, the pure-SRC scaling coefficient is
\begin{equation} \label{eq:pure-src-a2}
 a_2^{\mathrm{SRC}p}
 =\frac{a_2^p(^{12}\mathrm C)}{f_C}
 =22.45\pm0.85 \,.
\end{equation}
Inserting this result into Eq.\,\eqref{eq:emc-src-linear} yields
\begin{equation} \label{eq:pure-src-emc-slope}
 \left|\frac{\dd R_{\mathrm{EMC}}^{\mathrm{SRC}p}}{\dd x}\right|
 =1.94\pm0.12 \,.
\end{equation}
The stated error combines the fitted slope and the reported scaling coefficient of carbon $a_2^p(^{12}\mathrm C)$, with the proton fraction $f_C=0.20$ held fixed.  
It does not cover the model assumption that the measured linear relation extends to a hypothetical system made only of SRC protons.
\begin{figure*}[t]
    \centering
    \begin{minipage}[t]{0.48\linewidth}
        \centering
        \includegraphics[width=\linewidth]
            {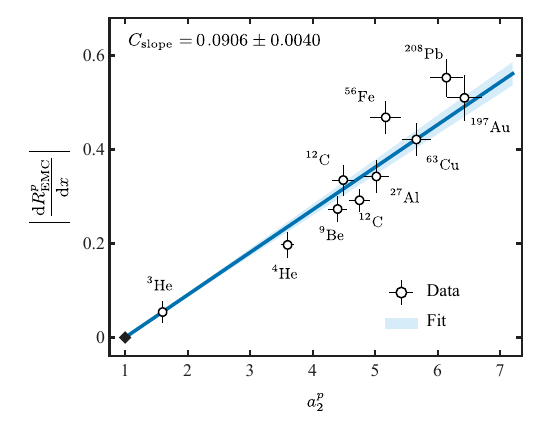}
        \par\vspace{2pt}
        \noindent
        \makebox[\linewidth][l]{%
            \hspace*{0.585\linewidth}%
            \makebox[0pt][c]{(a)}%
        }
    \end{minipage}%
    \hfill
    \begin{minipage}[t]{0.48\linewidth}
        \centering
        \includegraphics[width=\linewidth]
            {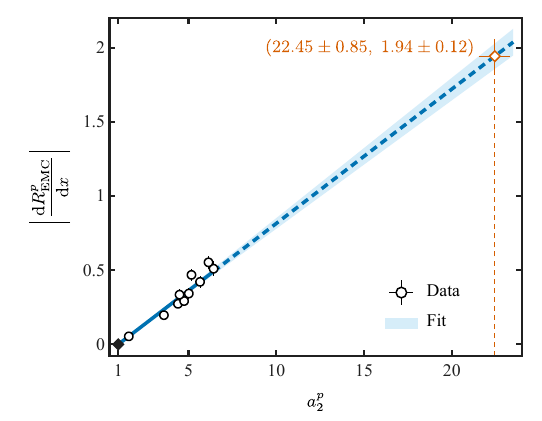}
        \par\vspace{2pt}
        \noindent
        \makebox[\linewidth][l]{%
            \hspace*{0.585\linewidth}%
            \makebox[0pt][c]{(b)}%
        }
    \end{minipage}

    \caption{Reconstruction of the pure-SRC EMC slope.
    Panel (a) shows the measured per-proton EMC-SRC linear
    correlation and the deuteron-anchored fit of
    Eq.\,\eqref{eq:emc-src-linear}.
    Panel (b) extends the same relation to $a_2^{\mathrm{SRC}p}=a_2^p(^{12}\mathrm{C})/f_C$.
    The solid line covers the measured region and the dashed
    line denotes the extrapolation.
    The displayed band propagates the fitted-slope uncertainty;
    it does not represent an uncertainty on the extrapolation model.}
    \label{fig:emc-src}
\end{figure*}

Fig.\,\ref{fig:emc-src} illustrates the extrapolation of the measured EMC–SRC relation to the pure-SRC limit. Using the measured SRC fraction in carbon as a normalization input, the assumed linear relation gives an estimate of the EMC slope for the pure-SRC proton ensemble. This external input helps constrain the rescaling freedom of the inclusive nPDF fit.

For comparison, we use the nCTEQ structure function ratio shown in Fig.\,2 of Ref.\,\cite{Denniston2024SRCnPDF} (the so-called baseSRC result). In the approximately linear EMC region, the curve gives
\begin{align} \label{eq:ncteq-pure-src-points}
 R^{\mathrm{SRC}p}_{\mathrm{nCTEQ}}(0.30)&\simeq0.852 \,,&
 R^{\mathrm{SRC}p}_{\mathrm{nCTEQ}}(0.35)&\simeq0.751 \,,\nonumber\\
 R^{\mathrm{SRC}p}_{\mathrm{nCTEQ}}(0.55)&\simeq0.304 \,,&
 R^{\mathrm{SRC}p}_{\mathrm{nCTEQ}}(0.60)&\simeq0.307 \,.
\end{align}
These values give a secant slope of approximately $-1.82$ over $0.30\leq x\leq0.60$ and $-2.22$ over $0.35\leq x\leq0.55$. A straight line fit over the broader interval gives approximately $-2.02$. We therefore characterize the magnitude of the slope as
\begin{equation} \label{eq:ncteq-pure-src-slope}
 \left|\frac{\dd R_{\mathrm{nCTEQ}}^{\mathrm{SRC}p}}{\dd x}\right|
 =2.02\pm0.20 \,.
\end{equation}

The slopes obtained in Eqs.\,\eqref{eq:pure-src-emc-slope} and \eqref{eq:ncteq-pure-src-slope} agree remarkably well. This agreement provides a nontrivial test because the former is reconstructed from the measured EMC-SRC linear correlation, whereas the latter is obtained from the pure-SRC structure function ratio extracted in the nCTEQ global fit. The two independent methods therefore give a consistent magnitude for the EMC modification associated with SRC protons, supporting the quark modification assigned to the SRC component by nCTEQ. This interpretation remains subject mainly to the measured value of carbon SRC fraction and the linear extrapolation to the pure-SRC limit.

The VMC comparison tests the nuclear dependence of the fitted SRC abundance, while the EMC–SRC relation provides an independent check of the quark modification in the EMC region. Together with the external SRC normalization, these results constrain the rescaling freedom of the two-component description. We next examine whether the same abundance coefficients describe the nuclear dependence of gluon modifications.

\section{Gluon Nuclear Modifications and the Rank-One Test}
\label{sec:gluon}
The quark sector gives the clearest link between the EMC effect and SRC observables, but a full partonic picture must also address gluons.
Gluons carry a large part of the nucleon momentum.
They also control scale evolution and many hard processes in nuclei. 
The short distances and large nucleon virtualities associated with SRC configurations may also modify the gluon distribution.
Recent works have studied linear quark and gluon EMC-SRC relations from this viewpoint \cite{Wang:2024cpx,Xu:2019wso,Wang:2024ikx,Hu2026LinearSRCEMC}. 
Here, we use a conventional nuclear PDF analysis given by EPPS21 to investigate whether the same abundance coefficients also organize the nuclear gluon modification.

\subsection{Reconstruction from EPPS21}
\label{sec:epps21-inversion}
EPPS21 determines nuclear modifications relative to the CT18A free proton baseline \cite{Hou:2019efy}. 
Its global data set includes fixed-target and LHC measurements.
Dijet and open-heavy-flavor data improve the gluon constraint.
For a proton bound in nucleus $A$, define
\begin{equation} \label{eq:epps21-ratio}
 \RgPA(x,Q^2)
 \equiv\frac{f_g^{p/A}(x,Q^2)}{f_{g,\mathrm{free}}^p(x,Q^2)} \,.
\end{equation}
Assuming isospin symmetry for the gluon modification in SRC pairs, i.e., $R_g^{\mathrm{SRC}\,p}=R_g^{\mathrm{SRC}\,n}\equiv R_g^{\mathrm{SRC}\,N}$, we will use $R_g^{\mathrm{SRC}\,N}$ to denote the common proton and neutron ratio below. The two-component form in Eq.\,\eqref{eq:two-component-proton} then gives
\begin{eqnarray} \label{eq:gluon-mixture}
 && \RgPA(x,Q^2) = 1 + \CpA\left(\RgSRC(x,Q^2)-1\right) \,,\nn\\
 && \RgSRC(x,Q^2) \equiv \frac{f_g^{\mathrm{SRC}N}(x,Q^2)}{f_{g,\mathrm{free}}^p(x,Q^2)} \,.
\end{eqnarray}
Here \(N\) denotes a proton or neutron, and \(f_g^{\mathrm{SRC}N}\) is the gluon PDF assigned to a nucleon in an SRC configuration.
Therefore,
\begin{equation} \label{eq:gluon-inversion}
 \RgSRC(x,Q^2) = 1+\frac{\RgPA(x,Q^2)-1}{\CpA} \,.
\end{equation}
Eq.\,\eqref{eq:gluon-inversion} makes the role of $\RgSRC$ clear.
The EPPS21 bound proton ratio differs from unity by only a few percent because it represents an average over all protons in the nucleus. 
If this modification is assigned only to the SRC fraction, $\CpA\simeq0.15$-$0.28$, its size must be divided by this fraction and therefore becomes much larger.
This division also magnifies any difference among nuclei. 
The agreement of the reconstructed SRC ratios is therefore a nontrivial result and is more restrictive than the apparent similarity of the original EPPS21 curves.
The free proton baseline in Eq.\,\eqref{eq:epps21-ratio} is the CT18A distribution used by EPPS21. 
We preserve that convention when constructing the absolute gluon distribution in Eq.\,\eqref{eq:gluon-mixture}.

The only parameter imported from nCTEQ analysis is $C_p^A$. 
Tab.\,\ref{tab:gluon-abundance} lists the coefficients used for the six nuclei, which are obtained from Eq.\,\eqref{eq:cpa-log-fit}. 
The reconstructed SRC gluon ratio relies on the nCTEQ two-component assumption.
EPPS21 alone determines the gluon distribution of an average bound proton and does not separate protons in SRC pairs from the remaining protons in the nucleus.

\begin{table}[tb]
\caption{Fitted nCTEQ coefficients used as SRC proton abundances in the EPPS21 reconstruction. The displayed entries are percentages.}
\label{tab:gluon-abundance}
\begin{ruledtabular}
\begin{tabular}{cccccc}
Nucleus & \(A\) & \(\CpA\) & Nucleus & \(A\) & \(\CpA\)\\[2pt]\hline\\[-5pt]
\(^{12}\mathrm C\) & 12 & 14.72 & \(^{64}\mathrm{Cu}\) & 64 & 22.51\\
\(^{27}\mathrm{Al}\) & 27 & 18.49 & \(^{197}\mathrm{Au}\) & 197 & 27.74\\
\(^{56}\mathrm{Fe}\) & 56 & 21.89 & \(^{208}\mathrm{Pb}\) & 208 & 27.99\\
\end{tabular}
\end{ruledtabular}
\end{table}

\begin{figure*}
\includegraphics[width=0.95\textwidth]{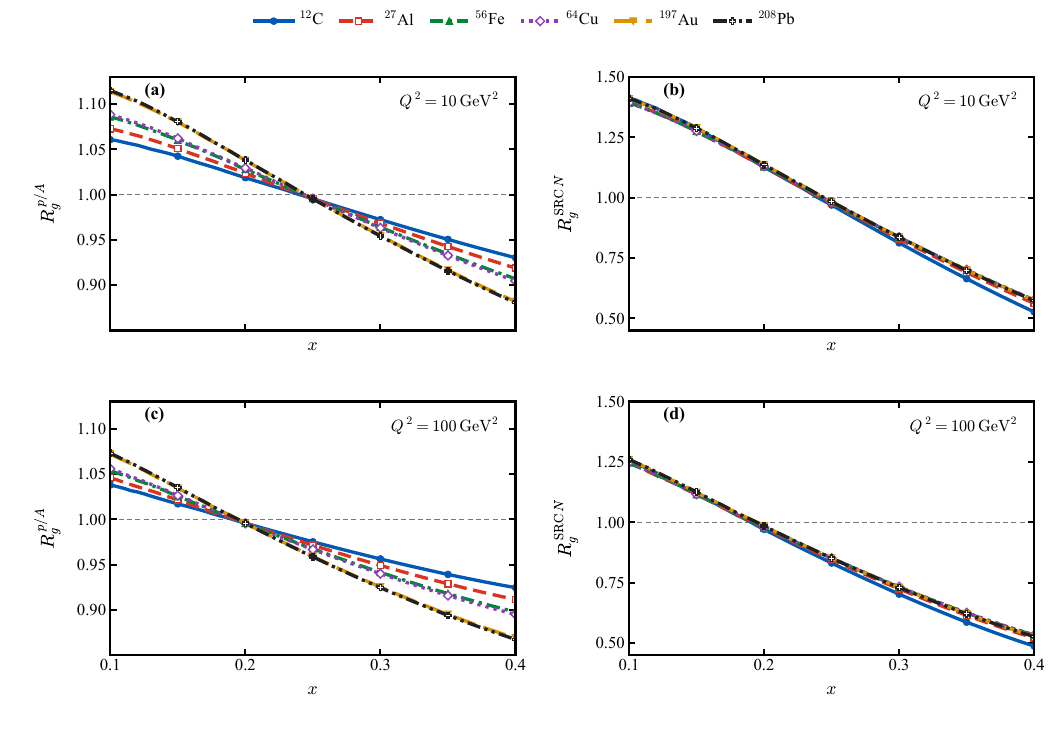}
\caption{EPPS21 central gluon modifications before and after the SRC reconstruction. The left panels show the bound proton ratios $R_g^{p/A}$; the right panels show the reconstructed SRC-nucleon ratios obtained from Eq.\,\eqref{eq:gluon-inversion}. The upper and lower rows correspond to $Q^2=10$ and $100\,\mathrm{GeV}^2$, respectively.}
\label{fig:gluon-collapse}
\end{figure*}

We focus on $0.1\leq x\leq0.4$, where the reconstructed ratios display an EMC-like transition from enhancement to suppression. 
The EPPS21 central grids are evaluated independently at $Q^2=10$ and $100\,\mathrm{GeV}^2$. 
Fig.\,\ref{fig:gluon-collapse} shows the main result of the reconstruction. 
After rescaling, the different $R_g^{p/A}$ central curves from carbon to lead collapse onto an approximately common $R_g^{\mathrm{SRC}\,N}$ at each scale. 
The reconstruction enhances the deviation from unity because the nuclear modification is assigned to a minority component. 
The original EPPS21 gluon ratios show clear differences across nuclei, whereas the reconstructed SRC gluon ratios fall close to a common curve.
This agreement provides a meaningful consistency test: the EPPS21 ratio and the nCTEQ coefficient are determined separately for each nucleus, and their combination yields a nearly universal result.

\subsection{Universality, scale evolution, and rank-one structure}
\label{sec:rank-one}

\begin{table*}[t]
\caption{Reconstructed SRC-nucleon gluon distributions at $Q^2=10\,\mathrm{GeV}^2$. Each $x$ group lists the ratio $R_g^{\mathrm{SRC}\,N}$ and the distribution $f_g^{\mathrm{SRC}\,N}$.}
\label{tab:gluon-q10}
\begin{ruledtabular}
\begin{tabular}{cccccccccc}
\multirow{2}{*}{Nucleus} & \multirow{2}{*}{\(\CpA\) (\(\%\))}& \multicolumn{2}{c}{\(x=0.1\)} & \multicolumn{2}{c}{\(x=0.2\)} & \multicolumn{2}{c}{\(x=0.3\)} & \multicolumn{2}{c}{\(x=0.4\)}\\
& & \(R_g^{\mathrm{SRC}\,N}\) & \(f_g^{\mathrm{SRC}\,N}\) & \(R_g^{\mathrm{SRC}\,N}\) & \(f_g^{\mathrm{SRC}\,N}\) & \(R_g^{\mathrm{SRC}\,N}\) & \(f_g^{\mathrm{SRC}\,N}\) & \(R_g^{\mathrm{SRC}\,N}\) & \(f_g^{\mathrm{SRC}\,N}\)\\[2pt]\hline\\[-5pt]
\(^{12}\mathrm C\)  & 14.72 & 1.41 & 18.21 & 1.13 & 2.94 & 0.81 & 0.69 & 0.53 & 0.17\\
\(^{27}\mathrm{Al}\) & 18.49 & 1.39 & 17.97 & 1.13 & 2.94 & 0.83 & 0.71 & 0.56 & 0.18\\
\(^{56}\mathrm{Fe}\) & 21.89 & 1.39 & 17.95 & 1.13 & 2.94 & 0.84 & 0.71 & 0.58 & 0.18\\
\(^{64}\mathrm{Cu}\) & 22.51 & 1.39 & 17.95 & 1.13 & 2.94 & 0.84 & 0.71 & 0.58 & 0.18\\
\(^{197}\mathrm{Au}\)& 27.74 & 1.41 & 18.16 & 1.14 & 2.96 & 0.84 & 0.71 & 0.57 & 0.18\\
\(^{208}\mathrm{Pb}\)& 27.99 & 1.41 & 18.19 & 1.14 & 2.96 & 0.84 & 0.71 & 0.57 & 0.18\\
\end{tabular}
\end{ruledtabular}
\end{table*}

\begin{table*}[t]
\caption{Reconstructed SRC-nucleon gluon distributions at $Q^2=100\,\mathrm{GeV}^2$, with the same notation as Tab.\,\ref{tab:gluon-q10}.}
\label{tab:gluon-q100}
\begin{ruledtabular}
\begin{tabular}{cccccccccc}
\multirow{2}{*}{Nucleus} & \multirow{2}{*}{\(\CpA\) (\(\%\))}& \multicolumn{2}{c}{\(x=0.1\)} & \multicolumn{2}{c}{\(x=0.2\)} & \multicolumn{2}{c}{\(x=0.3\)} & \multicolumn{2}{c}{\(x=0.4\)}\\
 &  & \(R_g^{\mathrm{SRC}\,N}\) & \(f_g^{\mathrm{SRC}\,N}\) & \(R_g^{\mathrm{SRC}\,N}\) & \(f_g^{\mathrm{SRC}\,N}\) & \(R_g^{\mathrm{SRC}\,N}\) & \(f_g^{\mathrm{SRC}\,N}\) & \(R_g^{\mathrm{SRC}\,N}\) & \(f_g^{\mathrm{SRC}\,N}\)\\[2pt]\hline\\[-5pt]
\(^{12}\mathrm C\)  & 14.72 & 1.26 & 14.15 & 0.97 & 1.93 & 0.70 & 0.40 & 0.49 & 0.09\\
\(^{27}\mathrm{Al}\) & 18.49 & 1.25 & 14.05 & 0.98 & 1.94 & 0.72 & 0.41 & 0.52 & 0.09\\
\(^{56}\mathrm{Fe}\) & 21.89 & 1.25 & 14.03 & 0.98 & 1.95 & 0.73 & 0.42 & 0.53 & 0.10\\
\(^{64}\mathrm{Cu}\) & 22.51 & 1.25 & 14.05 & 0.98 & 1.95 & 0.73 & 0.42 & 0.53 & 0.10\\
\(^{197}\mathrm{Au}\)& 27.74 & 1.26 & 14.19 & 0.98 & 1.96 & 0.73 & 0.41 & 0.53 & 0.10\\
\(^{208}\mathrm{Pb}\)& 27.99 & 1.26 & 14.19 & 0.98 & 1.96 & 0.73 & 0.41 & 0.53 & 0.10\\
\end{tabular}
\end{ruledtabular}
\end{table*}

Tables~\ref{tab:gluon-q10} and~\ref{tab:gluon-q100} list the reconstructed central values of $R_g^{{\rm SRC}\,N}$ and $f_g^{{\rm SRC}\,N}$ for each nucleus at four values of $x$. 
At $Q^2=10\,\mathrm{GeV}^2$, the reconstructed SRC ratio lies above unity at $x=0.1$ and $0.2$, and below unity at $x=0.3$ and $0.4$. 
At $Q^2=100\,\mathrm{GeV}^2$, the small-$x$ enhancement is weaker, and the value at $x=0.2$ lies slightly below unity. 
The reconstructed ratio therefore exhibits the expected scale dependence. 
Here, universality means that $\RgSRC$ is independent of the nuclear species at fixed $(x,Q^2)$, as
illustrated in Fig.\,\ref{fig:gluon-collapse}.

We quantify the common $x$ dependence through the matrix of gluon nuclear modifications. The two-component ansatz predicts a rank-one matrix whose nuclear factor is proportional to the nCTEQ coefficients $C_p^A$. A generic rank-one matrix need not satisfy this additional condition. Let $\ell$ label the nuclei and $x_m\in\{0.1,0.2,0.3,0.4\}$.
We define the modification matrix as
\begin{equation}
\label{eq:nuclear-modification-matrix}
M_{\ell m}=R_g^{p/A_\ell}(x_m,Q^2)-1\,,
\end{equation}
then the two-component ansatz predicts
\begin{equation}
\label{eq:rank-one-hypothesis}
M_{\ell m}=c_\ell h_m\,,
\quad
c_\ell=C_p^{A_\ell}\,,
\quad
h_m=R_g^{\mathrm{SRC}N}(x_m,Q^2)-1\,.
\end{equation}
Thus, $M$ has rank one with its nuclear factor fixed to $c_\ell$.
Equivalently, the reconstructed matrix
\begin{equation}
\label{eq:src-ratio-matrix}
S_{\ell m}=1+\frac{M_{\ell m}}{c_\ell}
\end{equation}
has identical rows. 
A generic rank-one matrix guarantees only that its rows are proportional; it does not fix their relative normalization.
To quantify the rank-one structure, we use the singular-value decomposition $X=\sum_k\sigma_k u_k v_k^{\mathrm T}$, where $\sigma_1\geq\sigma_2\geq\cdots$.
The Eckart--Young theorem \cite{Eckart1936} states that
the leading term \(X_1=\sigma_1u_1v_1^{\mathrm T}\) is the best rank-one
approximation and satisfies the following relation:
\begin{equation}
\|X-X_1\|_F^2=\sum_{k\geq2}\sigma_k^2 \,,
\end{equation}
where $\|\cdot\|_F$ denotes the Frobenius norm.
It therefore motivates the rank-one fraction
\begin{equation}
 F_1(X)=\frac{\sigma_1^2}{\sum_k\sigma_k^2} \,,
\end{equation}
which is exactly the fraction of the squared Frobenius norm captured by the
best rank-one approximation. Equivalently,
\(1-F_1=\|X-X_1\|_F^2/\|X\|_F^2\) is the minimum relative squared residual
of any rank-one description. Thus \(F_1=1\) if and only if \(X\) is rank
one, while \(F_1\simeq1\) quantifies approximate rank-one structure. Because
the common unity baseline can
make \(F_1(S)\) artificially large, we also quote \(F_1(S-\mathbf{1})\),
where \(\mathbf{1}\) is the all-ones matrix. It captures \(99.96\%\) and
\(99.98\%\) of the squared norm of \(S-\mathbf{1}\) at the two scales,
showing that the reconstructed modification matrix is nearly rank one.

\begin{table}[tb]
\caption{Rank-one tests for the six nuclei and four tabulated $x$ points. The entries are percentages.}
\label{tab:rank-one}
\begin{ruledtabular}
\begin{tabular}{ccc}
\(Q^2\,(\mathrm{GeV}^2)\) & \(F_1(S)\) & \(F_1(S-\bm 1)\)\\[2pt]\hline\\[-5pt]
10  & \(99.99\) & \(99.96\)\\
100 & \(99.99\) & \(99.98\)\\
\end{tabular}
\end{ruledtabular}
\end{table}

QCD evolution provides an additional consistency check. 
If the gluon and quark-singlet modifications of every nucleus share the same $C_p^A$ factor at the input scale, DGLAP evolution acts identically on all nuclei. 
It changes the $x$ dependence and mixes quarks with gluons, but it does not generate a new nuclear dependence. 
Heavy-flavor threshold matching and convolution with common hard coefficients preserve this factorization.
The modification matrix therefore remains approximately rank one under evolution, even though its $x$ dependence evolves.
The similar rank-one fractions at the two scales indicate that the common nuclear pattern is stable under QCD evolution.

%We restrict the analysis to central values on purpose.  A full error study would need the correlations among EPPS21 nuclei and \(x\) bins, as well as the covariance of the nCTEQ abundance parameters.  Treating all points as independent would overstate the information.  Adding separate Hessian bands for each nucleus would also miss the covariance of the low-rank test.  In addition, the smooth \(A\) dependence built into EPPS21 can favor a low-dimensional pattern.  We therefore use both rank-one measures only to describe the central result.  A comparison with alternative nuclear directions requires a future covariance analysis.

We now summarize the results of the preceding two sections.
The VMC comparison supports the nuclear dependence of the SRC abundance, while the reconstructed pure-SRC EMC slope supports the magnitude of the quark modification.
In the gluon sector, dividing the EPPS21 modifications by $\CpA$ brings the results from carbon to lead close to a common $R_g^{\mathrm{SRC}\,N}$, and the corresponding modification matrix is nearly rank one.
These findings are consistent with a simple physical picture: the nucleus primarily determines the abundance of SRC configurations, while the partonic response of such a configuration is approximately universal across nuclei.
The explicit values of $\CpA$, and thus the magnitude assigned to the SRC modification, still depend on the functional form used in the nCTEQ fit. Nevertheless, the independent tests in Secs.\,\ref{sec:independent-tests} and \ref{sec:gluon} introduce information beyond the original fit, reduce the practical ambiguity caused by the rescaling freedom, and support the physical interpretation of the nCTEQ results, although without completely removing their parametrization dependence.

Inclusive data primarily constrain the combination $\CpA(\RgSRC-1)$. Testing its connection to SRCs at the event level requires probing gluons while tagging an SRC-rich configuration in the same event. This can be achieved through recoil-tagged charm-production in DIS, where a backward high-momentum proton constrains the state of the struck neutron and inclusive charm selects the gluon channel.

\section{Collinear Factorization for Recoil-Tagged Charm-Production DIS}
\label{sec:tagged-charm}

We now construct an experimental observable that selects both the nuclear configuration of SRC and a gluon sensitive hard process. The framework consists of three steps, as illustrated below
\begin{equation}{\label{eq:explain_framework_Factorization}}
\begin{aligned}
     W_{d,R}^{\mu\nu;c\bar c}
 &\xrightarrow[]{\rm leading \ power} C_g^{c\bar c}\otimes\Mtag_{g/d\to p_{\mathrm{rec}}} \\
&\xrightarrow[]{\rm PWIA}
 \Simp_d^{\mathrm{LF}}
 \left[C_g^{c\bar c}\otimes g_{n|\Gamma_R}\right] \,.
\end{aligned}
\end{equation}
We first define the exact recoil-tagged hadronic tensor.
We then apply leading-power QCD factorization and introduce a recoil-differential extended fracture function.
Finally, we impose PWIA on this fracture function and express it in terms of a deuteron light-front spectral factor and a conditional gluon distribution. 
Accessing this conditional gluon distribution is an important goal of this section.

\subsection{Tagged semi-inclusive DIS kinematics}
\label{sec:tagged-kinematics}
\begin{figure}[tb]
\centering
\includegraphics[width=0.95\columnwidth]{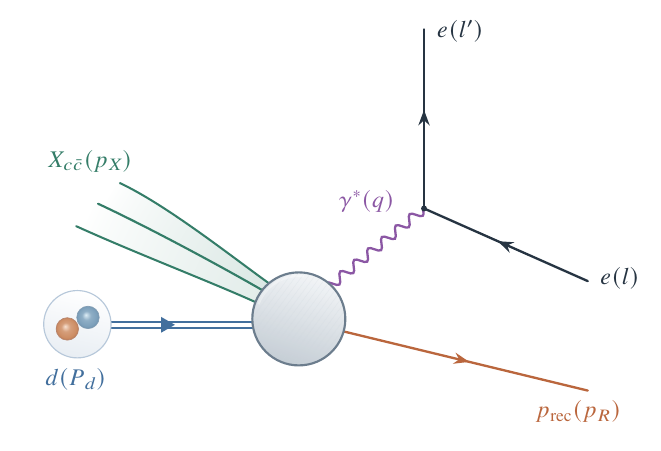}
\caption{Schematic diagram of recoil-tagged inclusive charm-production DIS on the deuteron. The measured recoil proton belongs to the target-fragmentation region, while the unobserved current region is summed inclusively subject to the presence of charm and anticharm.}
\label{fig:tagged-process}
\end{figure}

Specifically, we consider the scattering process
\begin{align}
 & e(l)+d(P_d) \to e(l')+p_{\mathrm{rec}}(p_R)+X_{c\bar c}(p_X) \,,
 \nonumber\\
 & \qquad\qquad q\equiv l-l' \,, \qquad Q^2\equiv-q^2>0 \,.
 \label{eq:tagged-process}
\end{align}
The process is illustrated in Fig.\,\ref{fig:tagged-process}.
We choose the detected recoil nucleon to be a proton because protons are experimentally easier to be detected than neutrons.
Its struck partner in the deuteron is therefore a neutron.
The state \(X_{c\bar c}\) denotes an inclusive sum over unobserved hadronic final states containing a produced \(c\bar c\) pair.
No exclusive \(D\bar D\) or charmonium state is selected.

The target-level DIS variables are
\begin{equation} \label{eq:deuteron-dis-variables}
 x_d\equiv\frac{Q^2}{2P_d\cdot q} \,,
 \qquad
 y\equiv\frac{P_d\cdot q}{P_d\cdot l} \,,
 \qquad
 x\equiv2x_d \,.
\end{equation}
The last equality defines the per-nucleon variable used to display the cross section, which corresponds to the effective Bjorken variable for scattering from a nucleon in the deuteron. We take light-front components $v^\pm=(v^0\pm v^3)/\sqrt{2}$. In the collinear frame, $P_d$ has a large positive plus component and the virtual photon points mainly along negative $z$ \cite{StrikmanWeiss2018}. 
The recoil variables are
\begin{equation} \label{eq:recoil-variables}
 \alpha_R\equiv\frac{2p_R^+}{P_d^+} \,,
 \qquad
 \Gamma_R\equiv(\alpha_R,p_{RT}) \,,
 \qquad
 p_{RT}\equiv|\bm p_{RT}| \,.
\end{equation}
Thus the recoil proton carries the plus-momentum fraction $\alpha_R/2$ of the deuteron.

In the deuteron rest frame, the definition becomes
\begin{equation} \label{eq:alpha-rest-frame-main}
 \alpha_R=\frac{2(E_R+p_R^z)}{M_d} \,,
 \qquad
 E_R=\sqrt{M_N^2+|\bm p_R|^2} \,.
\end{equation}
The sign of $p_R^z$ follows our choice of light-front axis.
Since the virtual photon points mainly along $-z$, a proton emitted backward relative to the photon has $p_R^z>0$ and generally $\alpha_R>1$. This convention is useful experimentally because the longitudinal recoil and its transverse momentum can be varied separately.

We integrate the recoil azimuth throughout the main text for convenience. Starting from the invariant one-particle phase-space of the recoil proton, the corresponding measures are
\begin{align} \label{eq:recoil-measure}
 \dd\Gamma_R^{\mathrm{full}}
 &\equiv
 \frac{\dd\alpha_R\,p_{RT}\dd p_{RT}\,\dd\phi_R}
 {2(2\pi)^3\alpha_R} \,,
 \nonumber\\
 \dd\Gamma_R
 &\equiv\int_0^{2\pi}\dd\Gamma_R^{\mathrm{full}}
 =\frac{\dd\alpha_R\,p_{RT}\dd p_{RT}}
 {2(2\pi)^2\alpha_R}
 =\frac{\dd\alpha_R\,\dd p_{RT}^2}
 {4(2\pi)^2\alpha_R} \,.
\end{align}
Quantities differential in $\dd\Gamma_R$ are densities at fixed $(\alpha_R,p_{RT})$, with $\phi_R$ integrated from $0$ to $2\pi$. App.\,\ref{app:kinematics} derives the related Jacobian and every phase-space factor.

At the target level, let $k$ be the gluon momentum entering the hard subgraph and define
\begin{equation} \label{eq:target-gluon-fraction}
 \xi_d\equiv\frac{k^+}{P_d^+} \,.
\end{equation}
The unobserved target remnant must retain nonnegative plus momentum.  Combining this exact upper bound with leading-power DIS support gives
\begin{equation}
 x_d\leq\xi_d\leq1-\frac{\alpha_R}{2} \,.
 \label{eq:target-support}
\end{equation}
For the production of a $c\bar c$ pair, define
\begin{equation} \label{eq:threshold-a}
 a\equiv1+\frac{4m_c^2}{Q^2} \,.
\end{equation}
When the incoming gluon is treated as collinear and massless, the threshold for producing a $c\bar c$ pair requires the stronger lower bound
\begin{equation} \label{eq:tagged-charm-support}
 \xi_d\geq a x_d \,,
 \qquad
 a x_d\leq1-\frac{\alpha_R}{2} \,.
\end{equation}
Fixing the recoil proton momentum also determines how much momentum remains for the hard scattering. The second condition states that this remaining momentum must be large enough to produce a $c\bar c$ pair. Eqs.\,\eqref{eq:target-gluon-fraction}-\eqref{eq:tagged-charm-support} need no struck neutron variable. We will introduce $x_n$ and $z_g$ only after making the PWIA assignment in Sec.\,\ref{sec:pwia}. A more detailed derivation of the above results is provided in App.\,\ref{app:kinematics}.

\subsection{Recoil-tagged hadronic tensor}
\label{sec:tagged-tensor}
With relativistic state normalization
\begin{equation} \label{eq:state-normalization}
 \langle p',\lambda'|p,\lambda\rangle
 =2E_p(2\pi)^3\delta^{(3)}(\bm p'-\bm p)
 \delta_{\lambda'\lambda} \,,
\end{equation}
it is convenient to abbreviate the current amplitude as
\begin{align} \label{eq:tagged-current-amplitude}
 {\cal J}_X^\mu(\lambda_R,\lambda_d)
 \equiv{}&\langle p_{\mathrm{rec}}(p_R,\lambda_R),X_{c\bar c}| J^\mu(0)|d(P_d,\lambda_d)\rangle \,.
\end{align}
Here $J^\mu$ is the electromagnetic quark current,
\begin{equation*}
 J^\mu(0) \equiv \sum_f e_f \, \bar\psi_f(0)\gamma^\mu\psi_f(0) \,,
\end{equation*}
where $f$ runs over the relevant quark flavors and $e_f$ is the quark electric charge. The fully recoil-differential hadronic-tensor density is then
\begin{align} \label{eq:tagged-hadronic-tensor}
 W_{d,R}^{\mu\nu;c\bar c}(P_d,q;p_R)
 \equiv{}&\frac{1}{3}\sum_{\lambda_d,\lambda_R}
 \sum_{X_{c\bar c}}\frac{(2\pi)^4}{4\pi}
 \nonumber\\
 &\times\delta^{(4)}(P_d+q-p_R-p_X)
 \nonumber\\
 &\times\bigl[{\cal J}_X^\mu(\lambda_R,\lambda_d)\bigr]^*
 {\cal J}_X^\nu(\lambda_R,\lambda_d) \,.
\end{align}
The factor $1/3$ averages the three deuteron helicities.  We sum the recoil proton spin and all unobserved states containing charm quark. Eq.~\eqref{eq:tagged-hadronic-tensor} defines the recoil-differential hadronic tensor without assuming PWIA. The four-dimensional delta function enforces momentum conservation, while the two current matrix elements form an amplitude and its complex conjugate.
The four dimensional delta function enforces momentum conservation. The two current matrix elements form an amplitude and its complex conjugate. The phase-space integrals hidden in the sum over $X_{c\bar c}$ make the current region inclusive. Electromagnetic current conservation gives $q_\mu W^{\mu\nu}=q_\nu W^{\mu\nu}=0$.

Experimentally, charm is identified through observables such as reconstructed $D$ mesons or displaced decay vertices. The measured yield must then be corrected for decay branching fractions, detector acceptance, efficiency, and resolution so that it represents the inclusive charm contribution to the structure function. The event definition should also remain unchanged by unresolved soft or collinear radiation, which is the meaning of an infrared-safe definition. On the theory side, we sum over all hadronic final states containing a produced $c\bar c$ pair and do not fix the momentum or hadronization of either heavy quark. The factorization formula therefore uses the standard inclusive heavy-flavor coefficient and does not require a charm fragmentation function.

The tensor used below is the azimuth-averaged density
\begin{equation} \label{eq:azimuth-averaged-tensor}
 W_{d,R}^{\mu\nu;c\bar c}(P_d,q;\Gamma_R)
 \equiv\frac{1}{2\pi}\int_0^{2\pi}\dd\phi_R\,
 W_{d,R}^{\mu\nu;c\bar c}(P_d,q;p_R) \,.
\end{equation}
At fixed recoil momentum, the tensor also contains $\cos\phi_R$ and $\cos2\phi_R$ terms \cite{StrikmanWeiss2018}. They vanish in the azimuthal average used here, so no dependence on $\phi_R$ will appear in the subsequent derivation.

In the one-photon exchange approximation and the DIS limit $4x_d^2M_d^2/Q^2\ll1$, contraction with the unpolarized leptonic tensor gives
\begin{align} \label{eq:tagged-cross-section}
 \frac{\dd\sigma^{c\bar c,\mathrm{tag}}}
 {\dd x\,\dd Q^2\,\dd\Gamma_R}
 ={}&\frac{2\pi\alpha_{\mathrm{em}}^2 Y_+}{xQ^4} \left[
\widetilde F_{2,d}^{c\bar c,\mathrm{tag}}
 -\frac{y^2}{Y_+}\widetilde F_{L,d}^{c\bar c,\mathrm{tag}}
 \right] \,,
 \nonumber\\
 Y_+\equiv{}&1+(1-y)^2 \,.
\end{align}
Every tilded structure function is a density with respect to \(\dd\Gamma_R\). 
We define the recoil-tagged charm reduced cross section by removing the known electromagnetic and leptonic factors,
\begin{align} \label{eq:tagged-reduced-cross-section}
 \widetilde\sigma_{\mathrm{red},d}^{c\bar c,\mathrm{tag}}
 &\equiv
 \frac{xQ^4}{2\pi\alpha_{\mathrm{em}}^2Y_+}
 \frac{\dd\sigma^{c\bar c,\mathrm{tag}}}
 {\dd x\,\dd Q^2\,\dd\Gamma_R}
 \nonumber\\
 &=\widetilde F_{2,d}^{c\bar c,\mathrm{tag}}
 -\frac{y^2}{Y_+}\widetilde F_{L,d}^{c\bar c,\mathrm{tag}} \,.
\end{align}
At moderate $y$, $\widetilde F_2$ dominates the reduced cross section. 
Separating $\widetilde F_2$ and $\widetilde F_L$ requires data at different beam energies or a joint fit with both functions.
For simplicity, we present only $\widetilde F_2$ in the rest of this paper. Similar factorization and PWIA matching steps also apply to $\widetilde F_L$. 
App.\,\ref{app:tensor} gives the details of related tensor contraction.

The reduced cross section remains differential in the recoil measure. 
Its normalization and dimension therefore differ from those of standard inclusive reduced cross section. 
The inclusive result is recovered only after summing over a complete set of final states.

\subsection{Extended fracture function}
\label{sec:extended-fracture}

An ordinary PDF gives the momentum fraction of the parton entering the hard scattering and sums over the target remnant. By contrast, a fracture function also conditions on a hadron observed in the target-fragmentation region \cite{TrentadueVeneziano1994}. In DIS, this region follows the incoming target and contains target-collinear remnants. It differs from the current-fragmentation region made by the struck parton. An extended fracture function leaves the observed hadron's transverse momentum or invariant momentum transfer unintegrated \cite{Grazzini1998CutVertices}. In this work, the recoil proton in Eq.\,\eqref{eq:tagged-process} is such a tag.

For a continuous recoil momentum, introduce the projector selected density
\begin{align} \label{eq:recoil-projector}
 \rho_R^{\mathrm{full}}(p_R)
 \equiv\frac{\dd\Pi_R}{\dd\Gamma_R^{\mathrm{full}}}
 &={}\sum_{\lambda_R,X}
 |p_{\mathrm{rec}}(p_R,\lambda_R),X\rangle
 \nonumber\\
 &\quad\times
 \langle p_{\mathrm{rec}}(p_R,\lambda_R),X| \,.
\end{align}
Here $\dd\Pi_R$ is the projector onto final states containing a recoil proton in the infinitesimal phase-space element $\dd\Gamma_R^{\mathrm{full}}$; consequently, $\rho_R^{\mathrm{full}}$ is the corresponding projector per unit recoil phase-space. Its azimuthal average is
\begin{equation} \label{eq:recoil-projector-average}
 \rho_R(\Gamma_R)
 \equiv\frac{1}{2\pi}\int_0^{2\pi}\dd\phi_R\,
 \rho_R^{\mathrm{full}}(p_R) \,.
\end{equation}
This object selects the kinematics of recoil proton, $\alpha_R$, and $p_{RT}$, sums its spin, and includes all unobserved target-collinear states. Because the recoil momentum is continuous, a state with one exact momentum has delta function normalization and does not define a finite projector by itself. Thus we here use the projector density $\rho_R(\Gamma_R)$. The product $\rho_R(\Gamma_R)\dd\Gamma_R$ selects recoil protons within an infinitesimal phase-space element, while integration over a finite experimental acceptance gives a well-defined projector. The factors of $2\pi$, $\alpha_R$, and transverse momentum in these expressions come from the recoil phase-space measure in Eq.\,\eqref{eq:recoil-measure}.

Let $n^\mu$ be a lightlike vector conjugate to the fast deuteron momentum, with $n^2=0$. We define the gauge-invariant, unpolarized gluon extended fracture function as
\begin{align} \label{eq:tagged-gluon-operator}
 &\Mtag_{g/d\to p_{\mathrm{rec}}}(\xi_d,\Gamma_R;\mu)
 \nonumber\\
 &\equiv
 -\frac{1}{3\xi_d P_d\cdot n}
 \sum_{\lambda_d}
 \int\frac{\dd\lambda}{2\pi}
 \nonumber\\
 &\quad\times e^{-i\lambda\xi_d P_d\cdot n}
 \langle d,\lambda_d|
 n_\alpha F_a^{\alpha\rho}(\lambda n)
 \nonumber\\
 &\quad\times
 {\cal W}^{(A)}_{n,ab}(\lambda,\infty)
 \rho_R(\Gamma_R)
 {\cal W}^{(A)}_{n,bc}(\infty,0)
 \nonumber\\
 &\quad\times
 n_\beta F_c^{\beta}{}_{\rho}(0)
 |d,\lambda_d\rangle \,.
\end{align}
The Fourier phase selects $k\cdot n=\xi_d P_d\cdot n$. Future-pointing adjoint Wilson lines join the two field-strength tensors at light-cone infinity, as required for DIS. The factor $1/3$ averages the deuteron spin. Eq.\,\eqref{eq:tagged-gluon-operator} extends the Berera-Soper cut correlator to a tagged recoil proton \cite{BereraSoper1996}. Its recoil density normalization follows Eqs.\,\eqref{eq:recoil-measure} and \eqref{eq:recoil-projector}.

The support can be read directly from the target-collinear momentum condition in Eq.\,(\ref{eq:tagged-charm-support})
\begin{equation*}
 \xi_d+\frac{\alpha_R}{2}\leq1 \,,
 \qquad
 x_d\leq\xi_d\leq1-\frac{\alpha_R}{2} \,.
 \label{eq:fracture-support}
\end{equation*}

It is worth stressing that the correlator in Eq.\,\eqref{eq:tagged-gluon-operator} is not a product of a PDF and a fragmentation function. It keeps the correlations between the active parton and the measured target remnant in one nonperturbative object. This is the main benefit of the fracture function description. The measured proton is not produced independently after the hard scattering. Its momentum is correlated with the parton taken from the deuteron, and this correlation survives at leading twist. Compared with an ordinary fracture function, the extended fracture function retains the dependence on the momentum transfer $t=(P_d - p_R)^2$. It therefore depends on the retained recoil kinematics, $\Gamma_R=(\alpha_R,p_{RT})$, and preserves information that would otherwise be lost upon integration over $t$.

\subsection{Leading-power QCD collinear factorization}
\label{sec:qcd-factorization}
When $Q^2$ is much larger than the fixed hadronic scales, the recoil condition changes the target-collinear matrix element but not the short-distance scattering.  The general DIS factorization theorem with a fixed target-fragmentation condition therefore applies \cite{Collins1998Diffractive}. At the first nonzero order in the three light flavor Fixed-Flavor-Number Scheme (FFNS), only photon-gluon fusion contributes. With a common factorization and renormalization scale $\mu_F=\mu_R=\mu$, the structure function is
\begin{align} \label{eq:tagged-factorization}
 &\frac{\widetilde F_{2,d}^{c\bar c,\mathrm{tag}}(x_d,Q^2;\Gamma_R)}{x_d}
 \nonumber\\
 ={}&\int_{a x_d}^{1-\alpha_R/2}\frac{\dd\xi_d}{\xi_d}
 \nonumber\\
 &\times
 C_{2,g}^{c\bar c}\!\left(\frac{x_d}{\xi_d},\frac{Q^2}{m_c^2},\frac{\mu^2}{m_c^2}\right)
 \nonumber\\
 &\times
 \Mtag_{g/d\to p_{\mathrm{rec}}}(\xi_d,\Gamma_R;\mu)
 +\order\!\left(\frac{\Ltag^2}{Q^2}\right) \,,
\end{align}
where $x_d/\xi_d$ is the partonic Bjorken ratio. The tagged hadronic scale is summarized by
\begin{equation} \label{eq:tagged-scale}
 \Ltag^2\sim
 \max\{M_d^2,M_N^2,p_{RT}^2,|t'|,\LQCD^2\} \,.
\end{equation}
The hard coefficient in Eq.\,\eqref{eq:tagged-factorization} is the massive inclusive DIS coefficient \cite{Gluck:1987uk,Laenen:1992zk,Laenen:1992zk,Laenen:1992xs,Riemersma1995Charm}. At leading power, all recoil dependence lies in the extended fracture function.

The factorization follows the usual DIS argument. The measured proton remains in the single target-collinear subgraph. Soft gluons joining this subgraph to the current jets enter the Wilson lines. The inclusive sum over unobserved current states gives the required unitarity cancellations. Unlike diffractive hadron-hadron scattering \cite{Collins:1992cv}, DIS has only one incoming colored collinear sector, namely, the target-collinear sector associated with the incoming deuteron. We therefore apply the general proof of Ref.\,\cite{Collins1998Diffractive}, rather than presenting a separate but essentially equivalent proof.

At this point, it is useful to clarify the distinction between the extended fracture function used here and the diffractive PDF discussed in the literature \cite{BereraSoper1996}. A diffractive PDF requires the incoming hadron to remain intact in a restricted channel. The recoil-tagged fracture function has no such condition. It is therefore the more general distribution for a detected proton from a deuteron.

We next discuss the scale evolution of $\Mtag$. At fixed $\Gamma_R$, renormalization acts on the light-cone operators and gives homogeneous DGLAP-type evolution \cite{Grazzini1998CutVertices}:
\begin{align} \label{eq:tagged-dglap}
 &\frac{\partial\Mtag_{i/d\to p_{\mathrm{rec}}}(\xi_d,\Gamma_R;\mu)}
 {\partial\ln\mu^2}
 \nonumber\\
 ={}&\sum_j\int_{\xi_d/(1-\alpha_R/2)}^1\frac{\dd u}{u}\,
 P_{i\leftarrow j}(u,\as)
 \nonumber\\
 &\times
 \Mtag_{j/d\to p_{\mathrm{rec}}}\!\left(\frac{\xi_d}{u},\Gamma_R;\mu\right) \,.
\end{align}
Here $P_{i\leftarrow j}$ are the usual splitting kernels, and $i,j$ label parton species \cite{Gribov:1972ri,Gribov:1972rt,Altarelli:1977zs,Dokshitzer:1977sg,Eskola:2002yc}. The lower limit keeps the parent parton within the momentum left after the recoil proton is observed. The evolution is homogeneous at fixed nonperturbative recoil.  If transverse momentum or $t$ is integrated into a perturbative region, an inhomogeneous source term can appear in the conventional integrated fracture function.

Eq.\,\eqref{eq:tagged-factorization} does not require PWIA. Final-state interactions (FSI) and other target-fragmentation effects are contained in $\Mtag$, which can enter different observables with the same recoil definition but different hard coefficients. We now use PWIA to relate this tagged correlator to the deuteron wave function and the struck-neutron distribution.

\subsection{PWIA matching}
\label{sec:pwia}
In light-front PWIA, the hard current acts on the neutron. The recoil proton is a passive on-shell spectator at the hard vertex.  We neglect rescattering and two-body hard currents. With spin labels suppressed, the deuteron light-front wave function is normalized as
\begin{equation} \label{eq:deuteron-lf-normalization}
 \int_0^2\frac{\dd\alpha_R}{2(2\pi)^3\alpha_R(2-\alpha_R)}
 \int\dd^2\bm p_{RT}\,
 |\Psi_d(\alpha_R,\bm p_{RT})|^2=1 \,.
\end{equation}
The corresponding proton-tagged light-front spectral factor is
\begin{equation} \label{eq:lf-spectral-factor}
 \Simp_d^{\mathrm{LF}}(\alpha_R,\bm p_{RT})
 \equiv\frac{|\Psi_d(\alpha_R,\bm p_{RT})|^2}{2-\alpha_R} \,.
\end{equation}
Within PWIA, the product $\Simp_d^{\mathrm{LF}}(\Gamma_R)\dd\Gamma_R$ gives the probability of finding the recoil proton within the infinitesimal momentum range specified by $\dd\Gamma_R$ \cite{StrikmanWeiss2018}. Therefore, $\Simp_d^{\mathrm{LF}}$ is a probability density rather than a probability by itself. It describes the proton-neutron momentum distribution in the deuteron but contains no information about the partonic structure of the active neutron. It should therefore not be identified with either a parton distribution or the exact tagged fracture function.

The normalization in Eq.\,\eqref{eq:deuteron-lf-normalization} treats both nucleons in the same way. The factors $\alpha_R$ and $2-\alpha_R$ give the plus-momentum fractions of the recoil proton and active neutron. Spin-dependent wave-function components are summed and averaged as in the unpolarized tensor.

The momentum left for the active neutron is
\begin{equation}\label{eq:active-neutron-momentum}
 p\equiv P_d-p_R \,,
 \quad
 p^+=\frac{2-\alpha_R}{2}P_d^+ \,,
 \quad
 \bm p_T=-\bm p_{RT} \,.
\end{equation}
Because the recoil momentum $p_R$ is measured, Eq.\,\eqref{eq:active-neutron-momentum} fixes the active neutron momentum in PWIA. Its invariant virtuality is fixed by the measured recoil momentum,
\begin{align} \label{eq:active-virtuality}
 & t\equiv p^2 = \frac{2-\alpha_R}{2}M_d^2
 -\frac{2p_{RT}^2}{\alpha_R} - \frac{2-\alpha_R}{\alpha_R}M_N^2 \,,\nonumber\\
 & t'\equiv t-M_N^2 = M_d^2-2M_dE_R \,.
\end{align}
Here $E_R=\sqrt{M_N^2+|\bm p_R|^2}$ is the energy of the recoil proton in the deuteron rest frame. Recoil kinematics fixes the invariant \(t=(P_d-p_R)^2\).  Its interpretation as the virtuality of an active neutron is part of PWIA.  App.\,\ref{app:kinematics} gives the derivation.

In PWIA, the active neutron has momentum $p=P_d-p_R$. We can therefore define the Bjorken variable and gluon momentum fraction relative to this neutron rather than to the deuteron,
\begin{align} \label{eq:active-neutron-variables}
 x_n&\equiv\frac{Q^2}{2p\cdot q}
 =\frac{x}{2-\alpha_R}
 \left[1+\order\!\left(
 \frac{M_N^2,p_{RT}^2,|t'|}{Q^2}\right)\right] \,,\nonumber\\
 z_g&\equiv\frac{k^+}{p^+}
 =\frac{\xi_d}{1-\alpha_R/2} \,.
\end{align}
The invariant definition of $x_n$ is exact. Its relation to $x/(2-\alpha_R)$ is a leading-power result. The definition of $z_g$ is exact after assigning $p=P_d-p_R$. These variables separate the measured neutron kinematics from the gluon fraction in the hard process.

Let $g_{n|\Gamma_R}(z_g,\mu)$ denote the gluon distribution of the active neutron conditional on the recoil proton kinematics. The gluon-channel matching ansatz is
\begin{align} \label{eq:pwia-matching}
 \Mtag_{g/d\to p_{\mathrm{rec}}}^{\mathrm{PWIA}}
 (\xi_d,\Gamma_R;\mu)
 ={}&\frac{\Simp_d^{\mathrm{LF}}(\Gamma_R)}
 {1-\alpha_R/2} \,\,
 g_{n|\Gamma_R}\Big(
 \frac{\xi_d}{1-\alpha_R/2},\mu \Big)
 \nonumber\\
 &\times\theta(\xi_d)
 \theta(1-\alpha_R/2-\xi_d) \,.
\end{align}
The spectral factor gives the weight of the nuclear configuration. The function $g_{n|\Gamma_R}$ is a parton density in the active neutron momentum fraction. The factor $1/(1-\alpha_R/2)$ is the density Jacobian from $\xi_d$ to $z_g$, as shown in Eq.\,(\ref{eq:active-neutron-variables}). Eq.\,\eqref{eq:pwia-matching} is a PWIA matching relation whose accuracy depends on FSI and two-body mechanisms.

We next change the integration variable in Eq.\,\eqref{eq:tagged-factorization} from $\xi_d$ to $z_g$. This change of variables requires the following Jacobian:
\begin{equation} \label{eq:pwia-density-jacobian}
 \dd z_g=\frac{\dd\xi_d}{1-\alpha_R/2} \,,
 \qquad
 \frac{\dd\xi_d}{\xi_d}=\frac{\dd z_g}{z_g} \,.
\end{equation}
At the first nonzero order in $\alpha_s$, charm-production proceeds through photon--gluon fusion,
\begin{equation} \label{eq:photon-gluon-fusion}
 \gamma^*(q)+g(k)\to c(p_1)+\bar c(p_2) \,.
\end{equation}
The corresponding hard coefficient in Eq.\,\eqref{eq:tagged-factorization} is $C_{2,g}^{c\bar c,\mathrm{LO}}=e_c^2 a_s(\mu)h_{2,g}^{(1)}$, where $e_c=2/3$ and $a_s=\alpha_s/(2\pi)$. Substituting Eq.\,\eqref{eq:pwia-matching} into Eq.\,\eqref{eq:tagged-factorization} gives our main formula:
\begin{equation}\label{eq:pwia-lo-master}
\begin{aligned}
 \widetilde F_{2,d}^{c\bar c,\mathrm{tag,PWIA}}
 &(x_n,Q^2;\Gamma_R)
 \\
 ={}&\Simp_d^{\mathrm{LF}}(\Gamma_R)e_c^2 a_s(\mu)x_n
 \int_{a x_n}^1\frac{\dd z_g}{z_g}\,
 \\
 &\times
 h_{2,g}^{(1)}\!\left(\frac{x_n}{z_g},\frac{m_c^2}{Q^2}\right)
 g_{n|\Gamma_R}(z_g,\mu)
 \\
 &+\order(a_s^2)+\order\!\left(\frac{\Ltag^2}{Q^2}\right) \,.
\end{aligned}
\end{equation}
Eq.\,\eqref{eq:pwia-lo-master} separates the recoil-dependent deuteron spectral factor from the hard photon–gluon coefficient and the conditional gluon distribution of the active neutron. The spectral factor selects the deuteron configuration. The massive coefficient probes it through photon-gluon fusion. The function $g_{n|\Gamma_R}$ contains the conditional gluon structure of the active neutron. App.\,\ref{app:coefficient} gives the detailed coupling convention, threshold, and massive coefficient of $h_{2,g}^{(1)}$.

The lower limit $a x_n$, with $a=1+4m_c^2/Q^2$, reflects the charm-production threshold. As $a x_n$ approaches unity, the convolution probes large $z_g$. Farther from threshold, it samples a broader range. Extracting the gluon distribution therefore requires measurements at several $x_n$ and $Q^2$, as in an ordinary PDF fit.

To connect the fixed-recoil distribution $g_{n|\Gamma_R}$ with an SRC-enriched sample, we select a finite recoil phase-space region ${\cal C}_{\mathrm{SRC}}$ using cuts on the recoil proton momentum and direction. We then average over this region with the light-front spectral weight and define
\begin{equation} \label{eq:acceptance-averaged-gluon}
 \overline g_n^{\mathrm{SRC},{\cal C}}(z_g,\mu)
 \equiv
 \frac{
 \displaystyle\int_{{\cal C}_{\mathrm{SRC}}}\dd\Gamma_R\,
 \Simp_d^{\mathrm{LF}}(\Gamma_R)
 g_{n|\Gamma_R}(z_g,\mu)}{
 \displaystyle\int_{{\cal C}_{\mathrm{SRC}}}\dd\Gamma_R\,
 \Simp_d^{\mathrm{LF}}(\Gamma_R)}
\end{equation}
and the corresponding ratio
\begin{equation} \label{eq:conditional-gluon-ratio}
 R_g^{\mathrm{SRC},{\cal C}}(z_g,\mu)
 \equiv\frac{\overline g_n^{\mathrm{SRC},{\cal C}}(z_g,\mu)}
 {g_n^{\mathrm{free}}(z_g,\mu)} \,.
\end{equation}
The superscript ${\cal C}$ specifies the recoil phase-space region over which the average is taken. Different recoil cuts can, in principle, select different neutron configurations and therefore give different gluon distributions. However, if ${\cal C}$ is genuinely enriched in SRC configurations and the gluon modification is universal within this region, the ratio defined in Eq.\,\eqref{eq:conditional-gluon-ratio} should remain approximately stable under reasonable variations of ${\cal C}$ and can be compared directly with the $R_g^{\mathrm{SRC}\,N}$ reconstructed in Sec.\,\ref{sec:gluon}.

\subsection{Discussion, corrections, and domain of validity}
\label{sec:tagged-validity}
The logical chain of the framework in this section is
\begin{equation} \label{eq:three-layer-chain}
 W_{d,R}^{\mu\nu;c\bar c}
 \longrightarrow C_g^{c\bar c}\otimes\Mtag_{g/d\to p_{\mathrm{rec}}} \longrightarrow
 \Simp_d^{\mathrm{LF}}
 \left[C_g^{c\bar c}\otimes g_{n|\Gamma_R}\right] \nn\,.
\end{equation}
The first object is an exact tagged tensor. The first arrow is a leading-power QCD statement, and the second is based on PWIA assumption. If this assumption fails, the extended fracture function remains the correct leading-twist object.  It simply loses the interpretation as a product of a spectral factor and a recoil-conditioned gluon distribution.

This separation has practical value. A problem with the Born kernel calls for a higher-order calculation. A problem with PWIA calls for a better nuclear reaction model. Neither affects the tensor definition of $W_{d,R}^{\mu\nu;c\bar c}$.

We first discuss the higher-order calculation. The first nonzero contribution comes from the photon-gluon fusion channel and provides a useful leading-order baseline. At order $a_s^2$, one must include virtual corrections and real-gluon emission in the photon-gluon channel, as well as subprocesses initiated by light quarks \cite{Laenen:1992zk,Laenen:1992xs,Riemersma1995Charm}. The theoretical uncertainty should also be examined by varying the scale $\mu$ and the charm-quark mass. These higher-order contributions modify the perturbative coefficient functions and introduce additional quark terms, but they do not change the definitions of the tagged correlators or the recoil variables.

We then discuss the PWIA assumption. A large active neutron virtuality does not by itself invalidate PWIA.  It is part of the selected high-momentum configuration and can appear in the \(\Gamma_R\) dependence of \(g_{n|\Gamma_R}\).  PWIA fails instead through rescattering, two-body currents, nonnucleonic components, or any process in which the proton is not a passive spectator.  Backward, low-\(p_{RT}\) kinematics can suppress FSI, and their remaining size can be tested through the recoil-angle dependence. In the future, a quantitative extraction of $g_{n|\Gamma_R}$ therefore requires a systematic theoretical framework to include the remaining FSI and their dependence on the recoil kinematics \cite{StrikmanWeiss2018,Hauenstein2024LAD}.

%Data can separate some of these mechanisms.  Two-body currents can change the link between recoil momentum and the active parton fraction.  The non-SRC background should vary when the lower momentum cut moves through the transition region.  Fits to several neighboring recoil bins are therefore better than one integrated ``SRC'' bin.  The final partonic result can still be quoted as an acceptance average.

The final point concerns the power correction. The power expansion requires
\begin{equation} \label{eq:factorization-hierarchy}
 Q^2 \,, \, Q^2+4m_c^2 \gg M_N^2\,, \, p_{RT}^2\,,\, |t'|\,,\,\LQCD^2 \,.
\end{equation}
The term \(\order(\Ltag^2/Q^2)\) in Eq.\,(\ref{eq:pwia-lo-master}) represents these target-mass, recoil, and higher-twist corrections. Their size should be estimated by using different high-momentum deuteron wave functions and by studying how the result changes with the recoil kinematics. The selected conditions must satisfy both the scale hierarchy in Eq.\,(\ref{eq:factorization-hierarchy}) and the charm-production condition $a x_n\leq1$. Increasing $Q^2$ reduces the power-suppressed corrections, but at a fixed beam energy it also leaves less phase space for inelastic scattering and lowers the number of charm events. The experimental cuts must therefore balance theoretical accuracy, the charm-event rate, recoil proton acceptance, and the ability to study the recoil kinematics dependence.

\subsection{Experimental observables}
\label{sec:tagged-observables}
The primary observable is inclusive charm-production in the current region, differential in $x,Q^2,\alpha_R$, and $p_{RT}$. It does not require a reconstructed $D\bar D$ pair or charmonium state. Our basic experimental observable is the tagged reduced cross section defined in Eq.\,\eqref{eq:tagged-reduced-cross-section}. It can be measured either at fixed recoil kinematics $\Gamma_R$ or after integration over a specified region of the recoil phase space:
\begin{equation} \label{eq:acceptance-cross-section}
 \sigma_{\mathrm{red},{\cal C}}^{c\bar c}
 \equiv\int_{\cal C}\dd\Gamma_R\,\,
 \widetilde\sigma_{\mathrm{red},d}^{c\bar c,\mathrm{tag}}
 (x,Q^2,y;\Gamma_R) \,.
\end{equation}
At moderate $y$, Eq.\,\eqref{eq:pwia-lo-master} supplies the leading expression because $\widetilde\sigma_{\mathrm{red}}\simeq\widetilde F_2$. Otherwise the corresponding longitudinal coefficient must be included. The acceptance ${\cal C}$ must be stated in measured variables. These include recoil momentum, polar angle, transverse momentum, and detector limits. Data and theory must use the same definition in Eq.\,\eqref{eq:acceptance-cross-section}.

The recoil cuts serve two different purposes: a large recoil momentum selects SRC-enriched configurations, while backward emission with small $p_{RT}$ suppresses FSI. A representative momentum selection
\begin{equation*} \label{eq:src-recoil-window}
 300\,\mathrm{MeV}\lesssim|\bm p_R|\lesssim600\,\mathrm{MeV}
\end{equation*}
selects more high momentum configurations. A proton emitted nearly opposite to the virtual photon, with small $p_{RT}$, also reduces rescattering. In our axis convention, the photon points mainly along $-z$. A backward proton therefore points mainly along $+z$ and usually has $\alpha_R>1$. The final kinematic selection must balance acceptance, statistics, SRC purity, and FSI.

%The momentum and angular cuts play different roles.  The momentum cut selects the high-relative-momentum part of the deuteron wave function.  The angular cut reduces rescattering between the current debris and the recoil proton.  The momentum cut alone can give an SRC-rich but distorted sample.  The angular cut alone can give a clean sample dominated by mean-field motion.  Their overlap defines the useful tagged region.

%Several control samples should accompany the SRC-rich region.  Lower recoil momenta test the normalization and the change from mean-field to correlated motion.  Sideways recoil angles increase FSI and constrain its model.  A wider \(p_{RT}\) cut tests stability against transverse recoil.  Inclusive tagged DIS in the same bins provides a high-statistics normalization and a quark-dominated comparison.

Here, we propose a useful companion observable which divides the charm tagged reduced cross section by inclusive tagged DIS in the same kinematic condition:
\begin{align} \label{eq:charm-to-dis-ratio}
 {\cal R}_{c\bar c/\mathrm{DIS}}
 (x,Q^2,y;\Gamma_R)
 \equiv
 \frac{\widetilde\sigma_{\mathrm{red},d}^{c\bar c,\mathrm{tag}}
 (x,Q^2,y;\Gamma_R)}
 {\widetilde\sigma_{\mathrm{red},d}^{\mathrm{DIS},\mathrm{tag}}
 (x,Q^2,y;\Gamma_R)} \,.
\end{align}
At moderate $y$ and in PWIA assumption, the common light-front spectral factor cancels,
\begin{equation} \label{eq:pwia-charm-to-dis-ratio}
 {\cal R}_{c\bar c/\mathrm{DIS}}^{\mathrm{PWIA}}
 \simeq
 \frac{F_{2,n}^{c\bar c}(x_n,Q^2|\Gamma_R)}
 {F_{2,n}^{\mathrm{DIS}}(x_n,Q^2|\Gamma_R)} \,.
\end{equation}
Because the numerator and denominator use the same recoil selection, their ratio reduces common uncertainties from luminosity, recoil proton efficiency, and the deuteron spectral factor. The charm numerator probes the conditional gluon distribution through photon–gluon fusion, while inclusive tagged DIS provides a high-statistics reference. Their ratio retains the partonic convolutions and hard coefficients of both processes.

Inclusive tagged DIS provides a control sample for validating the recoil treatment and constraining final-state interaction models. With these effects constrained, tagged charm data can be analyzed using the same recoil selection and the massive charm coefficient to extract the conditional gluon distribution $g_{n|\Gamma_R}$. Averaging this distribution over the SRC-enriched region gives $R_g^{\mathrm{SRC},{\cal C}}$, which can then be compared with the $R_g^{\mathrm{SRC}\,N}$ reconstructed in Sec.\,\ref{sec:gluon}. This comparison tests whether the gluon modification selected by recoil tagging is consistent with the scaling inferred from inclusive nPDFs.

\section{Summary and Outlook}
\label{sec:summary}
We have tested the SRC-based description of nuclear PDFs using nuclear-structure, quark-sector, and gluon-sector comparisons. High-momentum proton fractions from VMC calculations are consistent in size and nuclear dependence with the fitted nCTEQ coefficients $C_p^A$. Combining the measured EMC--SRC relation with an approximate $20\%$ SRC fraction in $^{12}\mathrm C$ gives a pure-SRC EMC slope of $1.94\pm0.12$, consistent with the nCTEQ estimate of $2.02\pm0.20$ under the stated assumptions.
	
Using the nCTEQ $C_p^A$ coefficients to rescale EPPS21 gluon modifications, we find an approximately common curve for six nuclei over $0.1\leq x\leq0.4$. Projection onto the $C_p^A$ direction captures about $99.9\%$ of the squared nuclear modification norm at both $Q^2=10$ and $100\,\mathrm{GeV}^2$. This supports the proposed scaling, although the smooth nuclear dependence of EPPS21 may also contribute to the observed pattern.
	
To test the SRC interpretation in selected nuclear configurations, we formulated recoil-tagged charm-production in deuteron DIS using an extended fracture function. Within PWIA, the tagged structure function relates the deuteron light-front spectral factor to a conditional gluon distribution of the active neutron through the massive photon--gluon coefficient. Comparing tagged charm-production with inclusive tagged DIS at the same recoil kinematics provides a way to constrain this distribution and compare it with the gluon modification inferred from inclusive nPDFs.
	
Such a comparison requires control of final-state interactions, two-body currents, the deuteron wave function, and non-SRC backgrounds. At the EIC, measurements across several SRC-enriched recoil selections could test the stability of the extracted gluon distribution and its consistency with inclusive SRC scaling.

%%%%%%%%%%%%%%%%%%%%%
\section*{Acknowledgements}
%%%%%%%%%%%%%%%%%%%%%
This work was supported in part by the National Natural Science Foundation of China under Grant Nos. 12675117, 12125503, 12335003, 12475098.
J.X was supported in part by the Key Laboratory for Particle Astrophysics and Cosmology, Ministry of Education (MoE). 
S.M.H and Y.Z were supported by the Fundamental Research Funds for the Central Universities under No. lzujbky-2023-stlt01.

\onecolumngrid
\appendix
\section{Light-Front Recoil Kinematics, Active Neutron Virtuality, and Support Region}
\label{app:kinematics}
This appendix gives details of the kinematics used in Sec.\,\ref{sec:tagged-kinematics}.

\subsection{Recoil measure and azimuth convention}
\label{app:recoil-measure}
In the deuteron rest frame,
\begin{equation} \label{eq:alpha-rest-frame}
 \alpha_R=\frac{2(E_R+p_R^z)}{M_d} \,,
 \qquad
 E_R=\sqrt{M_N^2+p_{RT}^2+(p_R^z)^2} \,.
\end{equation}
At fixed $p_{RT}$,
\begin{equation} \label{eq:alpha-jacobian}
 \frac{\partial\alpha_R}{\partial p_R^z}
 =\frac{2}{M_d}\left(1+\frac{p_R^z}{E_R}\right)
 =\frac{\alpha_R}{E_R} \,.
\end{equation}
It follows that
\begin{equation} \label{eq:longitudinal-jacobian}
 \frac{\dd p_R^z}{E_R}=\frac{\dd\alpha_R}{\alpha_R} \,.
\end{equation}
Using $\dd^3 \bm p_R=\dd p_R^z\dd^2 \bm p_{RT}$, we combine the longitudinal Jacobian in Eq.\,\eqref{eq:longitudinal-jacobian} with the transverse momentum element $\dd^2 \bm p_{RT}$ to obtain the recoil phase-space measure:
\begin{equation} \label{eq:invariant-recoil-measure}
 \frac{\dd^3\bm p_R}{2E_R(2\pi)^3}
 =\frac{\dd\alpha_R\,\dd^2\bm p_{RT}}
 {2(2\pi)^3\alpha_R} \,.
\end{equation}
Making use of $\dd^2\bm p_{RT}=p_{RT}\dd p_{RT}\dd\phi_R$ reproduces $\dd\Gamma_R^{\mathrm{full}}$ in Eq.\,\eqref{eq:recoil-measure}. Integration over $0\leq\phi_R<2\pi$ yields
\begin{equation} \label{eq:appendix-recoil-measure}
 \dd\Gamma_R
 =\frac{\dd\alpha_R\,p_{RT}\dd p_{RT}}
 {2(2\pi)^2\alpha_R}
 =\frac{\dd\alpha_R\,\dd p_{RT}^2}
 {4(2\pi)^2\alpha_R} \,.
\end{equation}
This is the azimuth-integrated measure $\dd\Gamma_R$ used throughout the main text to define recoil-differential tagged quantities and to perform the recoil integrations in Eqs.\,\eqref{eq:acceptance-averaged-gluon} and \eqref{eq:acceptance-cross-section}.

\subsection{Active neutron virtuality}
\label{app:active-virtuality}
Choose $P_{dT}=0$. In the light-front convention $v^\pm=(v^0\pm v^3)/\sqrt2$, the recoil on-shell condition is
\begin{equation} \label{eq:recoil-onshell}
 2p_R^+p_R^-=M_N^2+p_{RT}^2 \,.
\end{equation}
With $p=P_d-p_R$,
\begin{equation} \label{eq:active-components}
 p^+=\frac{2-\alpha_R}{2}P_d^+ \,,\qquad
 p^-=P_d^--p_R^- \,,\qquad
 \bm p_T=-\bm p_{RT} \,.
\end{equation}
Since $2P_d^+P_d^-=M_d^2$ and $p_R^-=(M_N^2+p_{RT}^2)/(2p_R^+)$, direct substitution gives
\begin{align} \label{eq:virtuality-derivation}
 p^2
 &=2p^+p^--p_{RT}^2
 \nonumber\\
 &=\frac{2-\alpha_R}{2}M_d^2
 -\frac{2-\alpha_R}{\alpha_R}(M_N^2+p_{RT}^2)
 -p_{RT}^2
 \nonumber\\
 &=\frac{2-\alpha_R}{2}M_d^2
 -\frac{2p_{RT}^2+(2-\alpha_R)M_N^2}{\alpha_R} \,.
\end{align}
This proves the first line of Eq.\,\eqref{eq:active-virtuality} without imposing an on-shell condition on the active momentum. The shifted virtuality follows by subtracting $M_N^2$.

The same result has a compact invariant form:
\begin{align} \label{eq:tprime-rest}
 t'&=(P_d-p_R)^2-M_N^2
 \nonumber\\
 &=M_d^2-2P_d\cdot p_R
 =M_d^2-2M_dE_R \,,
\end{align}
where the last equality holds in the deuteron rest frame. The measured recoil momentum fixes $t'$; it is not an independent integration variable. Physical recoil kinematics gives $t'<0$, and $|t'|$ generally grows with $|\bm p_R|$.

The weak binding expansion shows the physical scale. Write $M_d=2M_N-\epsilon_d$, where $\epsilon_d$ is the deuteron binding energy \cite{StrikmanWeiss2018}, and expand $E_R=M_N+|\bm p_R|^2/(2M_N)+\cdots$. This gives
\begin{equation} \label{eq:tprime-weak-binding}
 t'=-2M_N\epsilon_d-2|\bm p_R|^2
 +\order\!\left(\epsilon_d^2,
 \frac{\epsilon_d|\bm p_R|^2}{M_N},
 \frac{|\bm p_R|^4}{M_N^2}\right) \,.
\end{equation}
Above the small binding contribution, the virtuality grows quadratically with recoil momentum. A high-momentum recoil tag therefore also selects a highly virtual active neutron.

For comparison with light-front deuteron wave functions, one may introduce an internal momentum \(\bm k\) through the free two-nucleon invariant mass
\begin{equation} \label{eq:free-two-body-mass}
 M_0^2=\frac{4(M_N^2+p_{RT}^2)}{\alpha_R(2-\alpha_R)}
 =4(M_N^2+|\bm k|^2) \,,
\end{equation}
which gives
\begin{equation} \label{eq:internal-lf-momentum}
 k_T=p_{RT} \,,\qquad
 k^z=(\alpha_R-1)
 \sqrt{\frac{M_N^2+p_{RT}^2}{\alpha_R(2-\alpha_R)}} \,.
\end{equation}
Eqs.\,\eqref{eq:free-two-body-mass} and \eqref{eq:internal-lf-momentum} convert the measured recoil variables $(\alpha_R,p_{RT})$ into the internal relative momentum $\bm k$ used in the light-front deuteron wave function. Eq.\,\eqref{eq:free-two-body-mass} defines $|\bm k|$ through the invariant mass of a free proton-neutron pair, while Eq.\,\eqref{eq:internal-lf-momentum} gives its transverse and longitudinal components. This conversion is needed to evaluate $\Psi_d$ and to express an SRC condition, such as a cut on $|\bm k|$, in terms of the measured recoil variables. The momentum $\bm k$ is not measured directly and does not enter the exact tagged tensor or the QCD factorization formula. It is introduced only in PWIA to evaluate the deuteron spectral factor and define the SRC-selected region.

\subsection{Partonic support and the charm threshold}
\label{app:partonic-support}
The target-level quantities are
\begin{align} \label{eq:appendix-target-definitions}
 x_d&\equiv\frac{Q^2}{2P_d\cdot q} \,,
 \qquad x\equiv2x_d \,,
 \nonumber\\
 \alpha_R&\equiv\frac{2p_R^+}{P_d^+} \,,
 \qquad \xi_d\equiv\frac{k^+}{P_d^+} \,.
\end{align}
If the unobserved target-collinear remnant carries momentum \(r\), plus-momentum conservation gives
\begin{equation} \label{eq:plus-bookkeeping}
 \xi_d+\frac{\alpha_R}{2}+\frac{r^+}{P_d^+}=1 \,.
\end{equation}
The positivity $r^+\geq0$ therefore implies the exact upper bound
\begin{equation} \label{eq:exact-upper-support}
 \xi_d\leq1-\frac{\alpha_R}{2} \,.
\end{equation}

For a two-nucleon light-front configuration, the recoil proton must satisfy $0<\alpha_R<2$. The active neutron carries the remaining fraction $1-\alpha_R/2$ of the deuteron plus momentum. As $\alpha_R$ approaches 2, this remaining momentum approaches zero, leaving no allowed momentum for the gluon entering the hard scattering. Charm-production imposes the stronger condition $a x_d\leq1-\alpha_R/2$, which may exclude large values of $\alpha_R$ even before this endpoint is reached. The recoil selection must therefore be chosen together with ($x_d, Q^2$), and the charm-production threshold.

At leading power, the gluon entering the hard subgraph is collinear and its Bjorken ratio is
\begin{equation} \label{eq:partonic-bjorken}
 \widehat x\equiv\frac{Q^2}{2k\cdot q}
 =\frac{x_d}{\xi_d}
 \left[1+\order\!\left(\frac{\Ltag^2}{Q^2}\right)\right] \,.
\end{equation}
The condition $\widehat x\leq1$ gives $\xi_d\geq x_d$. Combining it with Eq.\,\eqref{eq:exact-upper-support} gives Eq.\,\eqref{eq:target-support}. The upper limit follows from exact target-collinear momentum conservation. The lower limit uses leading-power hard kinematics.

After the PWIA assignment, the active neutron has
\begin{equation} \label{eq:pwia-plus}
 p^+=\left(1-\frac{\alpha_R}{2}\right)P_d^+ \,.
\end{equation}
Therefore
\begin{equation} \label{eq:zg-exact}
 z_g\equiv\frac{k^+}{p^+}
 =\frac{\xi_d}{1-\alpha_R/2} \,,
 \qquad
 0<z_g<1 \,,
\end{equation}
and
\begin{equation} \label{eq:zg-jacobians}
 \frac{\partial z_g}{\partial\xi_d}
 =\frac{1}{1-\alpha_R/2} \,,
 \qquad
 \frac{\dd\xi_d}{\xi_d}
 =\frac{\dd z_g}{z_g} \,.
\end{equation}
The first relation in Eq.\,\eqref{eq:zg-jacobians} generates the explicit Jacobian in the PWIA matching, while the convolution measure remains unchanged.

The invariant definition of the active neutron Bjorken variable is
\begin{equation} \label{eq:xn-invariant}
 x_n\equiv\frac{Q^2}{2p\cdot q} \,.
\end{equation}
In the chosen collinear frame, $q^-$ is the leading photon component. Hence
\begin{align} \label{eq:pdotq-reduction}
 2p\cdot q
 &=2p^+q^-+2p^-q^+-2\bm p_T\cdot\bm q_T
 \nonumber\\
 &=(2-\alpha_R)P_d\cdot q
 \left[1+\order\!\left(
 \frac{M_N^2,p_{RT}^2,|t'|}{Q^2}\right)\right] \,,
\end{align}
which gives
\begin{equation} \label{eq:xn-leading-power}
 x_n=\frac{x}{2-\alpha_R}
 \left[1+\order\!\left(
 \frac{M_N^2,p_{RT}^2,|t'|}{Q^2}\right)\right] \,.
\end{equation}
Eq.\,\eqref{eq:xn-invariant} is exact after $p$ is assigned; by contrast, Eq.\,\eqref{eq:xn-leading-power} is not.

For a massless collinear gluon, we define
\begin{align} \label{eq:partonic-invariant-mass}
 \widehat W^2
 &\equiv(q+k)^2
 =-Q^2+2k\cdot q
 =Q^2\left(\frac{1}{\widehat x}-1\right) \,.
\end{align}
Producing two on-shell charm quarks requires $\widehat W^2\geq4m_c^2$, or $\widehat x\leq1/a$. At the target level,
\begin{equation} \label{eq:threshold-equivalences}
 \xi_d\geq a x_d \,,\qquad
 a x_d\leq1-\frac{\alpha_R}{2} \,,\qquad
 a x\leq2-\alpha_R \,.
\end{equation}
After the leading-power PWIA mapping, the same condition is $a x_n\leq1$. It requires the lower limit of the $z_g$ convolution in Eq.\,\eqref{eq:pwia-lo-master} to lie below its upper limit.

Table\,\ref{tab:notation} summarizes the kinematic variables used throughout this work.

\begin{table*}[t]
\caption{Notations used for kinematic variables. “Exact thereafter” means exact after the PWIA momentum assignment \(p=P_d-p_R\) has been made.}
\label{tab:notation}
\begin{ruledtabular}
\begin{tabular}{>{\raggedright\arraybackslash}p{0.26\textwidth}
                >{\centering\arraybackslash}p{0.35\textwidth}
                >{\raggedright\arraybackslash}p{0.25\textwidth}}
Quantity & Relation & Description \\[2pt]\hline\\[-5pt]
Deuteron Bjorken variable & \(x_d=Q^2/(2P_d\cdot q)\) & definition, exact \\[5pt]
Per-nucleon external variable & \(x=2x_d\) & definition, exact \\[5pt]
Lepton inelasticity & \(y=P_d\cdot q/(P_d\cdot l)\) & definition, exact \\[5pt]
Recoil coordinates & \(\Gamma_R=(\alpha_R,p_{RT})\) & definition, exact \\[5pt]
Recoil plus fraction & \(p_R^+/P_d^+=\alpha_R/2\) & definition, exact \\[5pt]
Invariant recoil measure & \(\dd\Gamma_R^{\mathrm{full}}=\dd\alpha_R\,\dd^2\bm{p}_{RT}/(2(2\pi)^3\alpha_R)\) & phase-space identity, exact \\[5pt]
Full-deuteron gluon fraction & \(\xi_d=k^+/P_d^+\) & definition, exact \\[5pt]
Assigned active momentum & \(p=P_d-p_R\) & PWIA assignment \\[5pt]
Active neutron plus momentum & \(p^+=(1-\alpha_R/2)P_d^+\) & PWIA assignment, exact thereafter \\
Active neutron gluon fraction & \(z_g=\xi_d/(1-\alpha_R/2)\) & definition, exact thereafter \\[5pt]
Active neutron Bjorken variable & \(x_n=Q^2/(2p\cdot q)\) & definition, exact \\[5pt]
Collinear relation & \(x_n\simeq x/(2-\alpha_R)\) & leading-power/PWIA approximation \\
Recoil invariant and virtuality & \(t=(P_d-p_R)^2=p^2,\quad t'=t-M_N^2\) & external invariant; PWIA interpretation \\
Charm-threshold factor & \(a=1+4m_c^2/Q^2\) & hard-kinematic definition, exact \\
Hard-kernel argument & \(z=x_n/z_g\) & leading-power/PWIA variable \\
\end{tabular}
\end{ruledtabular}
\end{table*}
\FloatBarrier

\section{From the Tensor Contraction to the Tagged Charm Cross Section}
\label{app:tensor}
This appendix connects the current-correlation definition in Eq.\,\eqref{eq:tagged-hadronic-tensor} with the reduced cross section used in the main text. It also fixes the recoil azimuth and $x_d$ to $x$ conventions.

For effectively massless, unpolarized electrons, the leptonic tensor is
\begin{equation} \label{eq:leptonic-tensor}
 L^{\mu\nu}(l,l')
 \equiv\frac{1}{2}\sum_{\lambda_e,\lambda_{e'}}
 j_e^\mu\bigl(j_e^\nu\bigr)^*
 =2\left(l^\mu l'^\nu+l^\nu l'^\mu
 -g^{\mu\nu}l\cdot l'\right) \,.
\end{equation}
The factor \(1/2\) averages the incoming-electron spin, while the final spin is summed.

The separation between the lepton and hadron tensors follows directly from the one-photon amplitude
\begin{equation} \label{eq:one-photon-amplitude}
 {\cal A}_{X}^{c\bar c}
 =\frac{e^2}{q^2}\,
 j_{e,\mu}(l',l) \,
 \langle p_{\mathrm{rec}},X_{c\bar c}|J^\mu(0)|d\rangle \,.
\end{equation}
The electromagnetic current $J^\mu$ contains quark charges. The two electromagnetic vertices give $e^2$, and the photon propagator gives $1/q^2$. We square Eq.\,\eqref{eq:one-photon-amplitude}, sum the unobserved charm states, and use the tensor definition. The result is the tensor-level differential cross section at fixed recoil momentum:
\begin{equation} \label{eq:tensor-cross-section}
 \dd\sigma^{c\bar c,\mathrm{tag}}
 =\frac{4\pi e^4}{4I_{ed}Q^4}
 L_{\mu\nu}W_{d,R}^{\mu\nu;c\bar c}\,
 \dd\Gamma_{e'}\,\dd\Gamma_R^{\mathrm{full}} \,,
\end{equation}
where
\begin{equation} \label{eq:electron-flux-phase-space}
 I_{ed}\equiv\sqrt{(l\cdot P_d)^2-m_e^2M_d^2} \,,
 \qquad
 \dd\Gamma_{e'}\equiv
 \frac{\dd^3\bm l'}{2E_{e'}(2\pi)^3} \,.
\end{equation}
%The factor \(Q^{-4}\) is the squared photon propagator.  The extra \(4\pi\) in Eq.~\eqref{eq:tensor-cross-section} compensates the conventional \(1/(4\pi)\) in the hadronic tensor.

For negligible electron mass, $I_{ed}=l\cdot P_d$. The lepton tensor contains no recoil information, and the hadron tensor contains no lepton variables. All recoil dynamics therefore stays in $W_{d,R}^{\mu\nu;c\bar c}$ before contraction.

Before averaging over the recoil azimuth, the unpolarized tagged tensor contains four scalar functions \cite{StrikmanWeiss2018}. Terms with the transverse recoil vector generate $\cos\phi_R$ and $\cos2\phi_R$ modulations. Their integrals vanish:
\begin{equation} \label{eq:azimuth-harmonics}
 \int_0^{2\pi}\dd\phi_R\cos\phi_R
 =\int_0^{2\pi}\dd\phi_R\cos2\phi_R=0 \,.
\end{equation}
To show explicitly which terms in the hadronic tensor generate the $\phi_R$ dependence, we define
\begin{align} \label{eq:tagged-transverse-basis}
 \widetilde P^\mu
 &\equiv P_d^\mu-q^\mu\frac{P_d\cdot q}{q^2} \,,
 \nonumber\\
 g_\perp^{\mu\nu}
 &\equiv g^{\mu\nu}-\frac{q^\mu q^\nu}{q^2}
 -\frac{\widetilde P^\mu\widetilde P^\nu}{\widetilde P^2} \,,
 \nonumber\\
 p_{R\perp}^\mu&\equiv g_\perp^{\mu\nu}p_{R,\nu} \,.
\end{align}
The symmetric, parity-even tensor can then be organized schematically as
\begin{align} \label{eq:four-function-decomposition}
 W^{\mu\nu}_{d,R}
 ={}& A_1 g_\perp^{\mu\nu}
 +A_2\widetilde P^\mu\widetilde P^\nu + A_3\widetilde P^{\{\mu}p_{R\perp}^{\nu\}}
 +A_4\left(
 p_{R\perp}^\mu p_{R\perp}^\nu
 -\frac{p_{R\perp}^2}{2}g_\perp^{\mu\nu}
 \right) \,,
\end{align}
where the scalar coefficients are defined so that every term has the same physical dimension as the hadronic tensor, and the braces denote symmetrization of the enclosed Lorentz indices. The third tensor structure produces a term proportional to $\cos\phi_R$, while the symmetric traceless fourth structure produces a term proportional to $\cos2\phi_R$. Both terms integrate to zero over $0\leq\phi_R<2\pi$. After this integration, only the two azimuth-independent structure functions remain, and the tensor can be written as
\begin{align}\label{eq:tagged-tensor-decomposition}
 W_{d,R}^{\mu\nu;c\bar c}
 ={}&
 \left(-g^{\mu\nu}+\frac{q^\mu q^\nu}{q^2}\right)
 \widetilde F_{1,d}^{c\bar c,\mathrm{tag}} + \frac{\widetilde P^\mu\widetilde P^\nu}{P_d\cdot q}
 \widetilde F_{2,d}^{c\bar c,\mathrm{tag}} \,,
\end{align}
where $\widetilde P^\mu$ is defined in Eq.\,\eqref{eq:tagged-transverse-basis} and satisfies $q\cdot\widetilde P=0$. The tilded structure functions in Eq.\,\eqref{eq:tagged-tensor-decomposition} are densities in the azimuth-integrated recoil phase-space. Here the $\widetilde F_{2,d}^{c\bar c,\mathrm{tag}}$ is then used in the factorization formula Eq.\,\eqref{eq:tagged-factorization}, as well as the PWIA master formula Eq.\,\eqref{eq:pwia-lo-master}. We note that the magnitude $p_{RT}$ remains in $\widetilde F_1$ and $\widetilde F_2$; only its direction is averaged. This matters when the cross section is binned in transverse recoil.

The longitudinal combination is defined by
\begin{equation} \label{eq:tagged-longitudinal}
 \widetilde F_{L,d}^{c\bar c,\mathrm{tag}}
 \equiv
 \left(1+\frac{4M_d^2x_d^2}{Q^2}\right)
 \widetilde F_{2,d}^{c\bar c,\mathrm{tag}}
 -2x_d\widetilde F_{1,d}^{c\bar c,\mathrm{tag}} \,.
\end{equation}
We also introduce the virtual photon polarization parameter
\begin{equation} \label{eq:epsilon}
 \epsilon\equiv
 \frac{1-y-x_d^2y^2M_d^2/Q^2}
 {1-y+y^2/2+x_d^2y^2M_d^2/Q^2} \,.
\end{equation}
We contract Eq.\,\eqref{eq:tagged-tensor-decomposition} with Eq.\,\eqref{eq:leptonic-tensor}, and then change the electron phase-space to $(x_d,Q^2)$ and integrate over the recoil azimuth. The result is
\begin{align} \label{eq:exact-polarization-cross-section}
 \frac{\dd\sigma^{c\bar c,\mathrm{tag}}}
 {\dd x_d\,\dd Q^2\,\dd\Gamma_R}
 ={}&\frac{2\pi\alpha_{\mathrm{em}}^2y^2}
 {x_dQ^4(1-\epsilon)}
 \left[
 \left( 1 + \frac{4 M_d^2 x_d^2}{Q^2} \right) \widetilde F_{2,d}^{c\bar c,\mathrm{tag}}
 -\Big( 1-\epsilon \Big)
 \widetilde F_{L,d}^{c\bar c,\mathrm{tag}}
 \right] \,.
\end{align}
Here $\dd\Gamma_R$ no longer contains $\dd\phi_R$.

In the DIS limit $4x_d^2M_d^2/Q^2\ll1$,
\begin{align} \label{eq:epsilon-dis-limit}
 \epsilon&\to\frac{2(1-y)}{Y_+} \,,
 \qquad
 1-\epsilon\to\frac{y^2}{Y_+} \,,
 \nonumber\\
 \frac{y^2}{1-\epsilon}&\to Y_+ \,,
 \qquad
 Y_+\equiv1+(1-y)^2 \,.
\end{align}
Since $x=2x_d$,
\begin{equation} \label{eq:x-jacobian}
 \dd x=2\dd x_d \,,
 \qquad
 \frac{\dd x}{x}=\frac{\dd x_d}{x_d} \,.
\end{equation}
Eqs.\,\eqref{eq:epsilon-dis-limit} and \eqref{eq:x-jacobian} then give Eq.\,\eqref{eq:tagged-cross-section}. Removing the known prefactor and $Y_+$ gives the reduced cross section in Eq.\,\eqref{eq:tagged-reduced-cross-section}.

\section{Perturbative Photon-Gluon Fusion Input}
\label{app:coefficient}
This appendix specifies the perturbative convention in the PWIA master formula. It also derives the charm threshold and gives the massive gluon coefficient used in Eq.\,\eqref{eq:pwia-lo-master}. Most of the derivation can be found in Ref.\,\cite{Riemersma1995Charm}. However, for completeness, we present it in this appendix.

\subsection{Scheme, scale, and normalization}
\label{app:coefficient-conventions}
The first nonzero subprocess we are interested in is
\begin{align} \label{eq:appendix-pgf}
 &\quad \gamma^*(q)+g(k) \to c(p_1)+\bar c(p_2) \,,
 \nonumber\\
 &\quad q^2=-Q^2\,, \qquad p_1^2=p_2^2=m_c^2 \,.
\end{align}
We set the factorization and renormalization scales equal $\mu_F=\mu_R=\mu$. The expansion parameter is
\begin{equation}
 a_s(\mu)\equiv\frac{\alpha_s(\mu)}{2\pi}\,,
 \qquad
 \alpha_s(\mu)\equiv\frac{g_s^2(\mu)}{4\pi}\,.
 \label{eq:as-convention}
\end{equation}

%The FFNS choice also defines the charm structure function.  There is no incoming charm PDF, and the heavy-quark mass regulates the final-state collinear region.  Far above \(m_c\), a variable-flavor scheme can reorganize logarithms of \(Q^2/m_c^2\).  We do not do so here.  Our baseline is meant to show the photon-gluon origin and the physical threshold directly.

We define the charm contribution to the structure function as
\begin{equation} \label{eq:coefficient-convention}
 C_{2,g}^{c\bar c,\mathrm{LO}}(z,\lambda;\mu)
 \equiv e_c^2 a_s(\mu)h_{2,g}^{(1)}(z,\lambda) \,,
 \qquad e_c=\frac{2}{3} \,.
\end{equation}
We separate \(e_c^2\), leaving \(h_{2,g}^{(1)}\) as a purely QCD and kinematic function. The superscript $(1)$ denotes the first power of $a_s$, not a one-loop amplitude. Photon-gluon fusion is a tree-level process whose structure function contribution begins at order $\alpha_s$.

\subsection{Hard variables and charm threshold}
\label{app:coefficient-threshold}
For the coefficient function only, define
\begin{equation} \label{eq:hard-kernel-variables}
 z\equiv\frac{x_n}{z_g} \,,
 \qquad
 \lambda\equiv\frac{m_c^2}{Q^2} \,,
 \qquad
 \widehat s=(q+k)^2=Q^2\frac{1-z}{z} \,.
\end{equation}
The speed of either heavy quark in the $c\bar c$ center-of-mass frame is
\begin{equation} \label{eq:beta}
 \beta\equiv
 \sqrt{1-\frac{4m_c^2}{\widehat s}}
 =\sqrt{1-\frac{4\lambda z}{1-z}} \,.
\end{equation}
Reality of $\beta$ ($\beta \geq 0$), or equivalently $\widehat s\geq4m_c^2$, gives
\begin{equation} \label{eq:coefficient-threshold}
 z\leq\frac{1}{1+4\lambda}\,,
 \qquad
 z_g\geq(1+4\lambda)x_n=a x_n \,.
\end{equation}
This is the origin of the lower integration limit in Eq.\,\eqref{eq:pwia-lo-master}. The threshold can also be checked before PWIA. The relations $z=x_n/z_g$, $z_g=\xi_d/(1-\alpha_R/2)$, and $x_n=x/(2-\alpha_R)$ give $z=x_d/\xi_d$ at leading power. Thus $z\leq1/a$ is the same as $\xi_d\geq a x_d$. Combining this with $\xi_d\leq1-\alpha_R/2$ gives $a x\leq2-\alpha_R$. All these derivations lead to the same boundary.

\subsection{Massive gluon coefficient and Insertion in the PWIA formula}
\label{app:massive-coefficient}
In the convention of Eq.\,\eqref{eq:coefficient-convention}, the Born coefficient is \cite{Riemersma1995Charm}
\begin{align} \label{eq:massive-gluon-coefficient}
 h_{2,g}^{(1)}(z,\lambda)
 =2T_F\Bigg\{&
 \left[
 z^2+(1-z)^2+4z\lambda(1-3z)-8\lambda^2z^2
 \right]
 \ln\frac{1+\beta}{1-\beta} + \beta\left[
 -1+8z(1-z)-4z\lambda(1-z)
 \right]\Bigg\}
 \nonumber\\
 &\times \theta \left(1-(1+4\lambda)z\right) \,,
 \qquad T_F=\frac{1}{2} \,.
\end{align}
The factor $T_F$ is fixed by $\mathrm{Tr}(T^aT^b)=T_F\delta^{ab}$. The logarithm and $\beta$ terms arise from the massive two-body angular integration. The step function enforces the threshold in Eq.\,\eqref{eq:coefficient-threshold}.

%Two limits provide useful checks.  At threshold, \(\beta\to0\) and \(\ln[(1+\beta)/(1-\beta)]\to2\beta\).  The coefficient then vanishes with the two-body phase-space.  At fixed \(z<1\) and \(\lambda\to0\), the logarithm becomes large.  This is the standard heavy-flavor collinear logarithm.  It cannot be removed by setting \(m_c=0\) in the FFNS formula.  The two limits check the sign of the logarithm and the position of the step function.

%The variables used in the original heavy-flavor calculation are related to ours by
%\begin{align}
% \xi_{\mathrm{Riem}}&=\frac{Q^2}{m_c^2}=\frac{1}{\lambda},
% \nonumber\\
% \eta&=\frac{\widehat s}{4m_c^2}-1
% =\frac{1-(1+4\lambda)z}{4\lambda z},
% \qquad
% \beta=\sqrt{\frac{\eta}{1+\eta}}.
% \label{eq:riemersma-mapping}
%\end{align}
%The source convention puts dimensional factors and a different normalization of \(\alpha_s\) outside the tabulated coefficient.  Equation~\eqref{eq:massive-gluon-coefficient} translates it to the conventions of Eqs.~\eqref{eq:coefficient-convention} and \eqref{eq:pwia-lo-master}.  The tabulated function cannot be used without this mapping.

%The translation involves three choices: the expansion parameter, the gluon-density normalization, and the definition of \(F_2^{c\bar c}\).  Compensating changes can make a numerical curve look reasonable while leaving the absolute tagged cross section wrong.  We therefore treat Eqs.~\eqref{eq:as-convention}, \eqref{eq:coefficient-convention}, and \eqref{eq:appendix-master-check} as one linked set of definitions.

With $z=x_n/z_g$, Eqs.\,\eqref{eq:coefficient-convention} and \eqref{eq:massive-gluon-coefficient} give
\begin{align} \label{eq:appendix-master-check}
 \widetilde F_{2,d}^{c\bar c,\mathrm{tag,PWIA+LO}}
 (x_n,Q^2;\Gamma_R)
 ={}&\Simp_d^{\mathrm{LF}}(\Gamma_R)
 e_c^2a_s(\mu)x_n \int_{a x_n}^{1}\frac{\dd z_g}{z_g}\,
 h_{2,g}^{(1)}\left(\frac{x_n}{z_g},\frac{m_c^2}{Q^2}\right)
 g_{n|\Gamma_R}(z_g,\mu)
 \nonumber\\*
 &+\order(a_s^2)
 +\order\!\left(\frac{\Ltag^2}{Q^2}\right) \,.
\end{align}
The overall factors, integration measure, limits, and gluon argument reproduce Eq.\,\eqref{eq:pwia-lo-master}. The hard coefficient is perturbative. The PWIA matching introduces the nonperturbative quantities $\Simp_d^{\mathrm{LF}}$ and $g_{n|\Gamma_R}$. The $\order(a_s^2)$ term is a perturbative correction, whereas $\order(\Ltag^2/Q^2)$ is a hadronic power correction.

At order $a_s^2$, the gluon channel receives virtual corrections to $\gamma^\ast g\to c\bar c$ and real radiation contributions from $\gamma^\ast g\to c\bar c g$, while the complete NLO charm structure function also contains light-quark-initiated subprocesses. These higher-order contributions do not change the recoil variables or the definition of the extended fracture function. They only modify the hard coefficients and extend the sum over partonic channels.

\twocolumngrid

\end{document}